\documentclass[acmsmall]{acmart}

\AtBeginDocument{%
  \providecommand\BibTeX{{%
    \normalfont B\kern-0.5em{\scshape i\kern-0.25em b}\kern-0.8em\TeX}}}

\usepackage{amsmath}
\usepackage{algorithmic}
\usepackage[linesnumbered,ruled,vlined]{algorithm2e}
\usepackage{soul}
\usepackage{float}
\usepackage{subfig}
\usepackage{color, colortbl}

\definecolor{myblue}{HTML}{9ECAE1}

\newcommand{\cmmnt}[1]{}

\begin{document}

\title{PACE: A Program Analysis Framework for Continuous Performance Prediction}

\author{Chidera Biringa}
\email{cbiringa@umassd.edu}
\orcid{0000-0001-5904-2764}
\affiliation{%
  \institution{University of Massachusetts Dartmouth}
  \streetaddress{285 Old Westport Rd, North Dartmouth, MA}
  \city{Dartmouth}
  \state{Massachusetts}
  \country{USA}
  \postcode{02747}
}

\author{G{\"o}khan Kul}
\email{gkul@umassd.edu}
\orcid{0000-0001-6467-1979}
\affiliation{%
  \institution{University of Massachusetts Dartmouth}
  \streetaddress{285 Old Westport Rd, North Dartmouth, MA}
  \city{Dartmouth}
  \state{Massachusetts}
  \country{USA}
  \postcode{02747}
}

\renewcommand{\shortauthors}{C. Biringa and G. Kul}

\begin{abstract}
Software development teams establish elaborate continuous integration pipelines containing automated test cases to accelerate the development process of software. Automated tests help to verify the correctness of code modifications decreasing the response time to changing requirements. However, when the software teams do not track the performance impact of pending modifications, they may need to spend considerable time refactoring existing code. This paper presents \texttt{PACE}, a program analysis framework that provides continuous feedback on the performance impact of pending code updates. We design performance microbenchmarks by mapping the execution time of functional test cases given a code update. We map microbenchmarks to code stylometry features and feed them to predictors for performance predictions. Our experiments achieved significant performance in predicting code performance, outperforming current state-of-the-art by 75\% on neural-represented code stylometry features.
\end{abstract}


\received{20 February 2007}
\received[revised]{12 March 2009}
\received[accepted]{5 June 2009}

\maketitle

\section{Introduction}
\label{sec:introduction}
Software development teams often adopt agile practices and implement automated testing and integration architecture to ensure that implemented software requirements work as anticipated. However, this approach often excludes the performance implications of specific methods and their aptitude to increase the overall software execution time. If ignored, the quality of user interaction with software functionalities may diminish due to poor performance, especially when automated test scripts do not maintain a record of adjustments made to software features, making it strenuous to identify the changes that cause performance issues.

We polled 34 software developers working at 28 organizations, varying from software companies to federal agencies with in-house software development units. Twenty-nine developers reported that their software development environments run on continuous integration (CI) pipelines. CI~\cite{citools} is a process in agile software development to automatically test, verify, and integrate local software updates to the remote or production version of the software. Pipelines facilitate continuous integration by serializing algorithms and methods to achieve the objective of continually merging local changes to the primary software. 100\% of these CI pipelines in the poll utilize automated tests. However, some software tests to measure performance are not easy to automate, straining the test team's ability to ensure the software continues to deliver the intended functionality in an acceptable time frame during periods when the software changes rapidly. Therefore, automation of performance tests is crucial to provide the ideal user experience. 

The same poll indicated that 88\% of these software developers believe it is necessary to automate performance-focused testing. To stress the importance of performance in software, Pinterest~\cite{pavic2019why} increased its traffic and sign-ups by 15\% by reducing its latency by 40\%. Mobify~\cite{pavic2019why} found that a 100ms decrease in homepage load time increased their annual revenue by \$380,000 and that a similar reduction in checkout page load time increased their revenue by \$530,000. Moreover, BBC~\cite{pavic2019why} discovered they lose 10\% traffic for every additional second their site takes to load. Therefore, companies started to invest in performance testing to enhance their economic competitiveness. 

Traditional software performance estimation methods are classified broadly under Model-based~\cite{cortellessa2000deriving, menasce2000method, kahkipuro2000uml, bernardi2002uml, de2000uml, lindemann2002performance} and Trace-based~\cite{andolfi2000deriving, aquilani2001performance, petriu2002software, woodside2001automated} categories. Model-based methods are commonly abstract and estimate software performance by developing models based on specific software, hardware resources, and the utilization of application features and functionalities. Some of these models include (i) queueing network-based~\cite{cortellessa2000deriving}, (ii) architectural pattern~\cite{menasce2000method}, (iii) performance-driven unified modeling language~\cite{kahkipuro2000uml}, (iv) process-algebra-driven~\cite{bernardi2002uml}, (v) simulation-driven~\cite{de2000uml}, and (vi) stochastic-driven~\cite{lindemann2002performance}. Trace-based methods are aided by an interactive outline of software activities for performance estimations. Performance is obtained and logged, and then these values are used to perform estimations on the equivalent software with dissimilar hardware environments. Some of these methods include (i) Regression and Filtering~\cite{lee2003run}, (ii) Queueing Networks-Labeled Transition Systems~\cite{aquilani2001performance}, (iii) Queueing Networks-Message Sequence Chart~\cite{andolfi2000deriving}, and (iv) Layered Queueing Network-Use Case Maps~\cite{petriu2002software}. 

Techniques and methodologies adopted in model and trace-based methods vary significantly, and their applicability is predominantly contingent on software type and hardware components. Furthermore, the methods require a high setup time and effort. For example, one of the earliest software performance engineering methods~\cite{williams2002five} based on Queueing Networks (QN) requires the implementation to begin from the earliest through the latest stages of the development process. It also needs the application of Execution Graphs and QN that replicate software behavior and hardware resources. On the other hand, we facilitate the performance prediction during software development by embedding our approach into Git with no additional code on the software being developed or conversely no preliminary information about the software communicated into the model. 

Our literature review showed that model and trace-based methods are commonly integrated into software design to detect performance degradation. However, the application of these models requires developers to be knowledgeable about the behavioral, communication, and architectural variables of the software~\cite{zhang2015dwarfcode}. Hence, significant changes to the software can potentially lead to discrepancies between the inferred and actual performance of the model.

This paper proposes \textbf{\texttt{PACE}}, a \textbf{\texttt{P}}rogram \textbf{\texttt{A}}nalysis Framework for \textbf{\texttt{C}}ontinuous Performance Pr\textbf{\texttt{E}}diction. \texttt{PACE} adopts a comprehensive domain-oriented feature engineering and machine learning (ML) --- a subset of artificial intelligence tasked with discovering insightful patterns from large amounts of data and algorithms~\cite{sarker2021machine} --- approaches to learn and make predictions on the execution times of automated test cases given an update to the software. \texttt{PACE} fragments distributed version-controlled software repositories into branches deployed to a testing framework, where embedded functional test cases are executed and corresponding test times logged. 

Concretely, the contributions of this paper are as follows: 
\begin{itemize}
\item We introduce a program analysis framework for continuous performance prediction by mapping code stylometry features to the execution test times.
\item We propose a feature engineering technique for software performance prediction.
\item We propose adopting the execution time of automated test cases as a microbenchmark for software performance.
\end{itemize}

\texttt{PACE} addresses the following research questions:

\noindent \textbf{Research Question 1 (RQ1):} Automated performance software testing in CI pipelines can be exceedingly time-consuming and require the continuous execution of large code segments. Traditional performance estimation methods such as stochastic process~\cite{lindemann2002performance}, process algebra~\cite{bernardi2002uml} and queuing networks~\cite{andolfi2000deriving, aquilani2001performance, petriu2002software} provide a means of achieving the same goal with fewer computational resource expenditure. However, it is a complex task, most of which falls on the shoulders of the development team. ML has been applied most recently in software performance prediction research efforts~\cite{meng2017mira, ramadan2021comparative, samoaa2022tep, zhou2019deeptle, liu2021using, didona2015enhancing, guo2018data, kaltenecker2020interplay, velez2021white} with varying degrees predictive accuracy. \textbf{Hence, what is the predictive prowess of software performance given a code update?} 

We answered this question by creating microbenchmarks of code performance through executing available functional test cases in the software ($\mathit{y}$), after which we processed to select and transform code performance features ($\mathbb{X}$). Finally, we feed the test times and vector features to a regression function ($\phi$) tasked with predicting the software performance given an update $\hat{y} = f_{\phi}(x^{\prime})$. We experimented with several regression algorithms and found that a k-nearest neighbors regressor model performs the best in predictive accuracy and latency. \texttt{PACE-kNN} is the best-performing model, obtaining a (0.0763 $\in$ $\mathbb{D}$ (first case study) and 0.7366 on $\mathbb{D}^{\prime}$)  (second case study) average error rate \texttt{($\mu$(RMMR))} on experimental datasets. \texttt{($\mu$(RMMR))} denotes the average result of \texttt{RMSE, MSE, MAE}, and \texttt{RMSLE}, see Section~\ref{subsec:eval_metrics} for the description of aforementioned metrics.

\noindent \textbf{Research Question 2 (RQ2): {What is the delta performance impact of commit at (cn-1) given a (cn) code update?}}

We answered this question by further analyzing the generated results from RQ1 to ascertain the delta performance of commits. $\mathbb{D}$ has two positive and a single negative performance impact. $\mathbb{D}^{\prime}$ has a total of twenty-four positive and negative performance impacts respectively.

\noindent \textbf{Research Question 2 (RQ3):} An indispensable feature of building \texttt{PACE's} predictive models is the selection of code features and its consequential numerical statistic and distribution semantic transformation. Given that the primary goal of this work is continuously predicting code performance, it is pertinent that we do not introduce a significant performance overhead during the feature extraction pipeline. \textbf{Hence, what is the cost of selecting and representing code stylometry features?}

We answered this question and the concern of cost in throughput and latency by timing the selection,  statistic \texttt{(SR)}, and neural representation \texttt{(NR)} of code-stylometry features. The average selection throughput of aforementioned features takes (\texttt{0.2297} $\in$ $\mathbb{D}$ and \texttt{23.3201} $\mathbb{D}^{\prime}$) seconds (\texttt{42} $\in$ $\mathbb{D}$  and \texttt{2543} $\in$ $\mathbb{D}^{\prime}$) $\mu$ number of source files for our experimental datasets (see Section~\ref{subsec:datasets}). Statistic representation \texttt{(SR)} throughput takes (0.0002 $\in$ $\mathbb{D}$, 0.1269 $\in$ $\mathbb{D}^{\prime}$) average seconds, while neural representation takes (0.0055 $\in$ $\mathbb{D}$, 0.0588 $\in$ $\mathbb{D}^{\prime}$) average seconds. \texttt{SR} has a lower representation throughput compared to \texttt{NR}, outperforming it by (27\texttt{x} $\in$ $\mathbb{D}$ and 10\texttt{x} $\in$ $\mathbb{D}^{\prime}$). However, there is no significant difference in predictive performance between them when fed to predictors for performance prediction. 

\noindent \textbf{Research Question 4 (RQ4): {How does \texttt{PACE} perform in comparison with the state-of-the-art approach?}}

We answered this question by comparing the predictive performance of \texttt{PACE} to \texttt{TEP-GNN}~\cite{samoaa2022tep} --- a deep learning approach towards performance prediction based on graphical convolutional neural networks. Our best predictor \texttt{PACE-kNN} attained a (0.0083, 0.0077) \texttt{MSE} error rate on the entire dataset \texttt{(H2, RDF4J, Dubbo, SystemDS)}, and (0.0101, 0.0147) on average for independent projects, outperforming \texttt{TEP-GNN} (0.0170) (0.0202) by (68\%, 75\%) and (66\%, 31\%) $\in$ \texttt{(SR, NR)} features respectively.

\texttt{\textbf{Paper Outline.}} We organized the rest of this paper as follows. Section~\ref{sec:motivating_bg} defines the problem, and Section~\ref{sec:background} details the necessary background for this work. Section~\ref {sec:pace} details the \texttt{PACE} methodology. We describe the experimental setup and evaluation of \texttt{PACE} in Section~\ref{sec:evaluation}. In Section~\ref{sec:result}, we comprehensively answer research questions. Section~\ref{sec:scope} describes the \texttt{PACE}'s scope and threats. In Section~\ref{sec:related_work}, we discuss research works done in the software performance prediction literature. Finally, we conclude in Section~\ref{sec:conclusion}.

\section{Problem Definition: Unoptimized Code}
\label{sec:motivating_bg}
\begin{figure}[ht]
\centering
\includegraphics[width=0.90\linewidth]{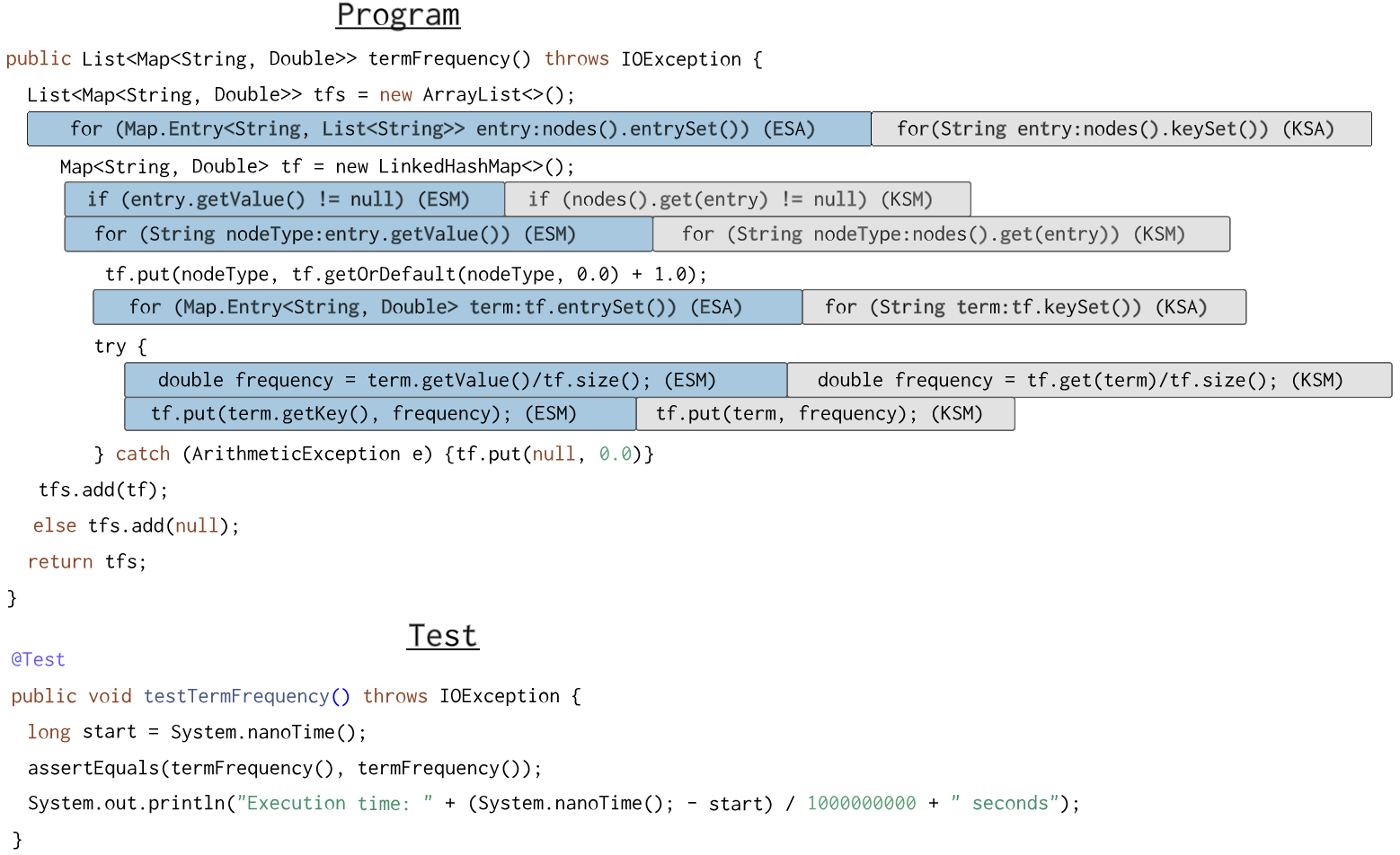}
\caption{Access and manipulation of a Linked-HashMap (LHM) using an Entry-set and Key-set. ESA and ESM denote Entry-set Access and Manipulation. KSA and KSM represent Key-set Access and Manipulation. ESA and KSA define a call to the sets (entry and key) view of the map, while ESM and KSM imply the modifications of elements in the map. Lines of code highlighted in blue and grey are LHM access and manipulation using  Entry-set and Key-set.}
\label{fig:problem} 
\end{figure}

We evaluate the performance impact of a snippet of source code by analyzing the time it takes to execute automated test cases. Execution time is a reliable metric on how users perceive software performance~\cite{sherief2014software}. In this paper, a code snippet or program is mediocre if it introduces a significant performance overhead to software and consequently skews the baseline resulting in an outlier performance. For example, Figure~\ref{fig:problem} is a code snippet executed on an Apple commodity laptop with an Apple M1 8-core CPU, 16-core Neural Engine, 8 GB RAM, and macOS Big Sur (version 11). 

The snippet is a real-world fragment of a Java program that calculates the term frequency of more than 2,000 elements in a linked hashmap \texttt{(LHM)}. An \texttt{LHM} is a data structure that combines a hash table and a linked list. It ensures predictable maintenance of elements in an iterable object~\cite{goodrich2014data}. The peripheral difference between the snippets is in how the linked hashmap is accessed and manipulated using an \texttt{entrySet()}, a set representation of \texttt{key,value} pairs, and \texttt{keySet()}, a set representation of \texttt{keys}. For both set representations, element changes in the map are reflected in the set and conversely.

The test execution of \texttt{(ESA, ESM)}, and \texttt{(KSA, KSM)} versions of the code snippet takes \textbf{$\sim$4 seconds} and \textbf{$\sim$4 minutes} respectively, indicating a \textbf{6000\%} increase in execution time. This time difference decreases by \textbf{$\sim$2 minutes}, a \textbf{3000\%}, given an increase in the number of elements from 2,000 to 200,000. The above-stated performance discrepancy might appear inconsequential for isolated occurrences. However, in scenarios where the low-performing program is called in multiple areas of software, it would increase the completion time of an operation. Furthermore, the \texttt{KSA} and \texttt{KSM} versions of the program, which are inefficient, are also syntactically correct and contain no apparent bugs, highlighting the problem's difficulty, such as employing erroneous data structure selection and manipulation and selecting $O(2^{n})$ time algorithms when there are viable polynomial options. Hence, in this paper, we approach this problem by mining and learning the features of low-performing programs and feeding those features to predictive models to detect their performance.

\section{Background}
\label{sec:background}
\subsection{Static Program Analysis}
\label{tinysec:sca}
 Static program analysis is the automated examination of programming code without its execution. Researchers have shown the importance of employing program analysis in solving code-dependant problems~\cite{louridas2006static} such as de-anonymizing of programmers~\cite{caliskan2015anonymizing}, authorship identification~\cite{dauber2018poster}, document representation and summarization~\cite{abuhamad2018large}, software defect prediction~\cite{li2017software}, plagiarism detection~\cite{lancaster2004comparison, cui2010code}, and software forensics~\cite{macdonell1999software}, and SQL query intent analysis~\cite{kul2020analysis} Program analysis using regular grammar is complex and incurs non-trivial performance overhead~\cite{karper2014efficient}. Hence, researchers commonly adopt an abstract syntax tree-enabled approach (AST) to program analysis.

An AST of a source code creates a semantic representation of the code~\cite{neamtiu2005understanding, zhang2019novel}. It captures a source code's implicit syntactic and lexical characteristics. ASTs are generated using a lexical analyzer that processes the source code and creates lexical tokens. The compiler passes these tokens to a syntactic analyzer and generates an AST. Finally, generated trees are converted to machine code and executed. 

\subsection{Microbenchmarking}
\label{tinysec:micro}
Software performance microbenchmarking is a standardized procedure to experimentally analyze the execution time of non-functional components of the software, such as code snippets. In contrast, spike, configuration, scalability, load, and stress tests assess the performance of the entire system using the benchmark characteristics of specific features and components~\cite{hooda2015software}. For instance, packages such as Java Microbenchmark Harness (JMH)~\cite{costa2019s} measure the performance of classes and functions in distributed and non-distributed computing environments. 

Although the above-stated programming language-based solution bridges a performance analysis gap in software development, it also introduces a bottleneck in implementation. For example, Costa \textit{et al.}~\cite{costa2019s} detailed the negative performance effect on benchmarking measures caused by bad coding practices by developers, which is akin to our motivating example. Even with formal computer science education, significant software documentation, and several standard online training courses, there is still a proliferation of unoptimized and bug-ridden code, especially in open-source software~\cite{lavallee2015good}. Hence, we guarantee a degree of program correctness by associating software performance with the execution time of automated test cases, which serves as the ground truth microbenchmarks for developing performance-based predictive models to predict code performance.

\section{PACE Framework}
\label{sec:pace}
\begin{figure}[ht]
\centering
\includegraphics[width=0.90\linewidth]{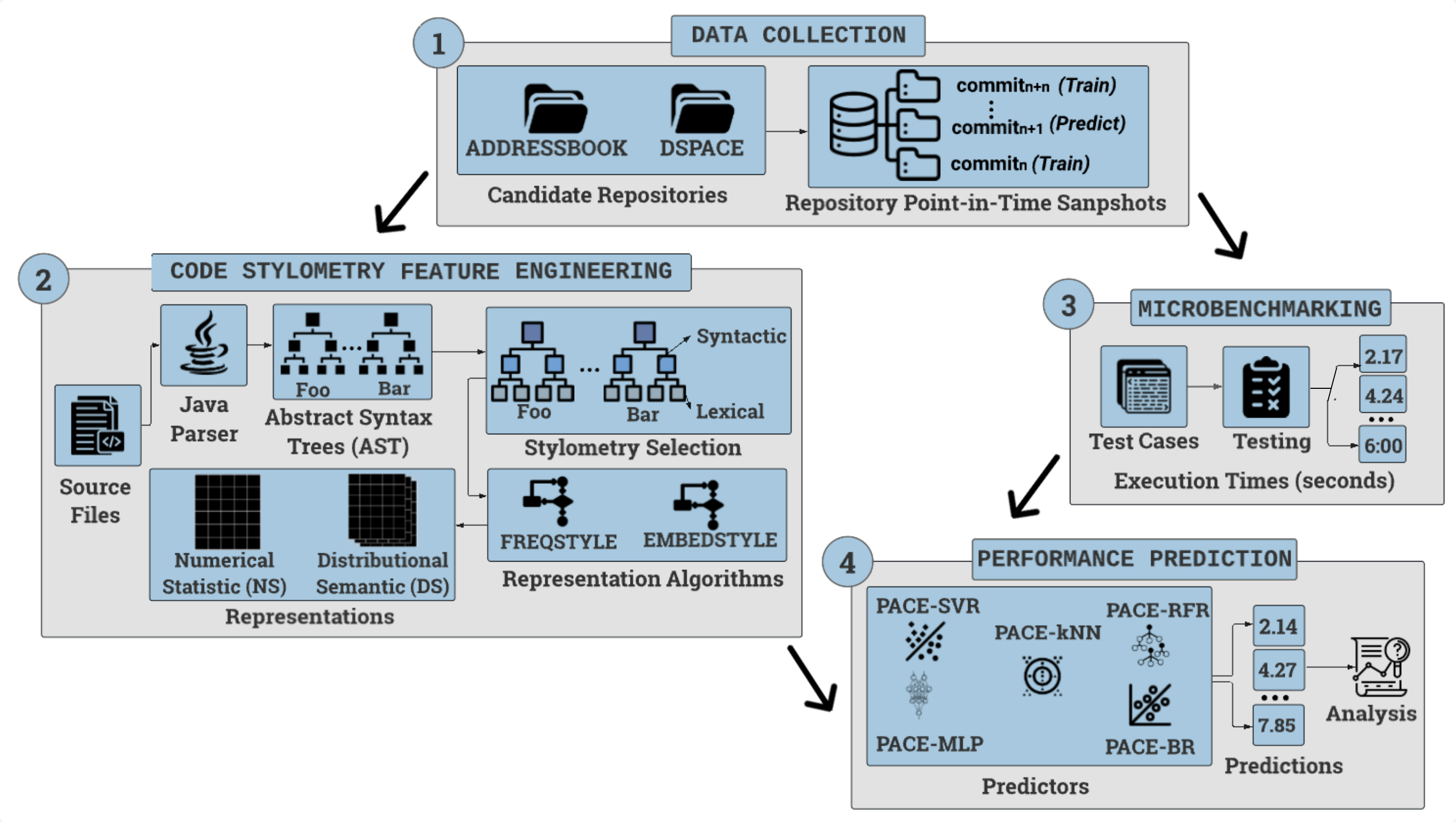}
\caption{Overview of PACE}
\label{fig:methodology}
\end{figure}

\texttt{PACE's} framework displayed in Figure~\ref{fig:methodology} is decomposed into four main phases: (i) in Phase 1 (see Section~\ref{subsec:data_collection}), we collect experimental data by selecting candidate repositories and simulating a version-controlled software development environment, (ii) Phase 2 (see Section~\ref{subsec:features}) details our code stylometry feature selection and representation methods. We selected syntactic and lexical features and transformed those features to real-valued vectors using proposed statistic and neural algorithms, (iii) in Phase 3 (see Section~\ref{subsec:automated_testing}), we extract the execution time of unit test cases using Travis CI and Maven Surefire, and (iv) Phase 4 (see Section~\ref{subsec:models}) describes the implementation of predictors for continuous performance predictions. We implemented several regression models and selected the best model for predicting code performance.
 
\subsection{Phase 1: Data Collection}
\label{subsec:data_collection}
The performance of predictive models is highly dependent on the quality of input data. How the data is collected is pertinent in filtering out bias and ensuring appropriate problem representation. This section details the strategies used in collecting experimental data for this study. In paragraph 1, we detail our data collection criteria. Paragraph 2 describes the decomposition of selected candidate repositories into \texttt{git branches} representing the current code state at time $n$ of a \texttt{git push}. Finally, paragraph 3 discusses \texttt{PACE's }continuous prediction strategy. 

\textbf{\texttt{Selecting Candidate Repositories.}} Instead of creating our experimental code repositories to evaluate our system, we selected existing popular candidate repositories. Candidate repositories used in our experiments were selected using characteristic filters that satisfy repository quality and popularity criteria as case studies. These filters are as follows: candidate repository must be (i) available and accessible on GitHub, (ii) contain compilable Java code, (iii) have test cases, (iv) have commits $\geq$ 24, and 12 non-contributor stars and forks. These parameters ensure that we achieve a representative collection of candidate repositories. In our poll among software developer participants, 97\% (33 out of 34) declared that their company requires developers to commit only compilable code, and 94\% (32 out of 34) expressed that they develop most unit and integration tests during requirements analysis. Therefore, our filters reflect expectations in a professional software project.

\textbf{\texttt{Repository Point-in-Time Snapshots (RPiTS).}} Post candidate repositories selection, we replicated the software development process by going back in time to freeze and extract the repository state given a commit$_{i}$. We reverse \texttt{(R)} repository history and extracted branches using the following steps:

\begin{itemize}
    \item [\texttt{R\textsubscript{1}}.] Forked the repository and initialized the cloned repository using the \texttt{git init} command.
     \item [\texttt{R\textsubscript{2}}.] Linked remote repository using the \texttt{git remote add origin <\textit{remote\_url}>} command.
     \item [\texttt{R\textsubscript{3}}.] Browsed through repository commit history using \texttt{git log}, and \texttt{git log --online} commands.
     \item [\texttt{R\textsubscript{4}}.] Created commit branches using the \texttt{git checkout -b <\textit{branch\_name}>} command.
     \item [\texttt{R\textsubscript{5}}.] Pushed repository's content to the newly created branch using the \texttt{git push -u origin <\textit{branch\_name}>} command.
     \item [\texttt{R\textsubscript{6}}.] Performed a hard reset on the repository branch using the \texttt{git reset --hard <\textit{commit\_hash}>} command.
\end{itemize}

\begin{figure}[ht]
\centering
\includegraphics[width=0.70\linewidth]{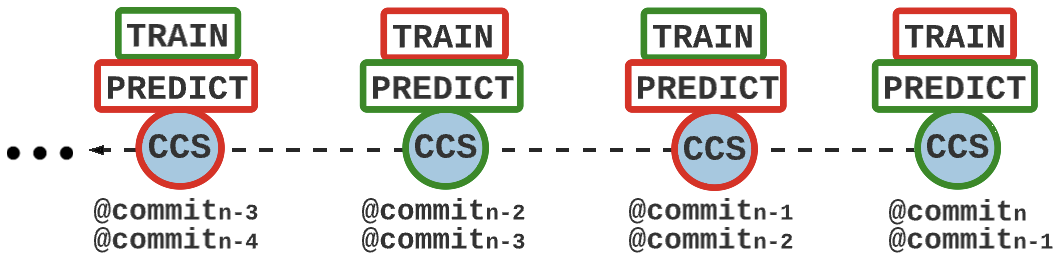}
\caption{Continuous Predictions. \texttt{CCS}: Current Code State. The input-features of a \texttt{CCS$_{n}$} given commit \texttt{$n$} (previous latest) are trained on a predictive model to predict the performance impact of \texttt{CCS$_{n-1}$} at commit \texttt{$n-1$} (current latest). Following that, we use \texttt{CCS$_{n-1}$} (previous latest (originally, current latest)) input-features in predicting \texttt{CCS$_{n-2}$} (current lastest) performance. We roll this process until \texttt{CCS$_{n-n}$}}.
\label{fig:rolling}
\end{figure}

\texttt{\textbf{Continuous Predictions and Deduplication.}} We propose continuous or rolling predictions in line with our goal of predicting the performance of the local software repository before being pushed to its remote counterpart. Encapsulated in the Figure~\ref{fig:rolling} displays base repository copy ($n$) is the repository's current code state (CCS), while ($n - 1$) is a local version consisting of an updated CCS in the queue to be pushed to remote. We feed  $n$ (training) to a regression function ($\phi$) to predict the performance of $n-1$ (testing). This process is repeated for ($n-n$). Furthermore, it eliminates duplicated observations. 

\subsection{Phase 2: Code Stylometry Feature Engineering}
Stylometry is the application of statistics to derive insights from the natural language writing style of a human~\cite{li2006fingerprint}. One of its earliest pieces of literature came in \textit{The Fredalist} papers~\cite{mosteller2012applied}. The authors, Mosteller and Wallace, compared unattributed corpora to those of their selected works. They applied statistical similarity methods to map anonymous corpus to probable authors. Motivated by this approach, Abbasi \textit{et al.}~\cite{abbasi2008writeprints} introduced a scalable stylometric analysis approach termed \textit{Writeprints}. The authors extracted salient lexical, content, syntactic, idiosyncratic, and structural features to create a robust feature set. Consequently, they fed these features to a Karhunen-Loeve Transform algorithm enabled by pattern disruptors to detect anonymized cyberspace entities. 

Code-stylometry (CStyle) is the stylometric analysis of a code's syntactic and lexical characteristics~\cite{tereszkowski2021towards}. In principle, cstyle is akin to stylometry. Contrastly, in stylometry, we analyze natural languages, while cstyle involves the analyses of programming languages. Researchers have successfully applied cstyle features in code analysis~\cite{sarnot2019snapcode} and security~\cite{caliskan2015anonymizing, kul2020analysis}. We leverage this knowledge to select syntactic and lexical features in Section~\ref{subsubsec:feature_select} and numerically transformed selected nodes in Section~\ref{subsubsec:feature_rep}.

\subsubsection{CStyle Feature Selection}
\label{subsubsec:feature_select}
Colin and Bernat~\cite{colin2002scope} presented \textit{scope-tree}, a methodology employed to dynamically ascertain the worst-case execution time (WCET) of a program. The authors analyzed the scope of a code using performance features such as declaration, expressions, loops, and conditionals. WCET's symbolical evaluation showed the utility of employing the aforementioned features in analyzing the worst-case execution time of a program. Thus, motivated by the \textit{scope-tree} approach, we selected syntactic and lexical features from source code files. Instead of treating characters and words in the corpus as tokenized entities, we selected syntactic and lexical features for numerical representations. 

Syntactic features are programming language dependent and provide context on a programmer's style-based preferential characteristics. These features vary in representation --- higher-level languages such as Python are more expressive and easier to comprehend, and lower-level languages such as C tend to be arcane with a higher comprehension bar~\cite{caliskan2015anonymizing}. Lexical features are semantic representations of language syntax that define the meaning of actions in a code~\cite{caliskan2015anonymizing}. Our targeted feature selection technique ensures a reduced corpus size (see Section~\ref{subsec:rq3}). Table~\ref{table:cstyle} details the taxonomy of selected features.

\label{subsec:features}
\begin{table}[ht]
\centering
\caption{A Taxonomy of Selected Code Stylometry Features.  Selected features represent five classes of nodes: statements, controls, expressions, invocations, and declarations.}
\resizebox{0.90\linewidth}{!}{
\begin{tabular}
{{ p{0.12\linewidth} p{0.55\linewidth} p{0.45\textwidth} }}
\toprule
\multicolumn{3}{c}{\textbf{Features: \{{\texttt{Statements, Controls, Expressions}}\} $\in$ Syntactic $\land$ \{{Invocations, Declarations}\} $\in$ Lexical}} \\
\toprule
\centering{\textbf{Class}} & \centering{\textbf{Types}} & \textbf{Brief Description} \\
\toprule
\centering{Statements} & \centering{IfStatement, WhileStatement, DoStatement, AssertStatement, SwitchStatement, ForStatement, ContinueStatement, ReturnStatement, ThrowStatement, SynchronizedStatement, TryStatement, BreakStatement, BlockStatement, BinaryOperation, CatchClause} & Dictates the behavior of a program under explicitly defined conditions \\
\midrule
\centering{Controls} & \centering{For, EnhancedFor} & Defines the repetition of instructions dependent on the satisfaction of requirements \\
\midrule
\centering{Expressions} & \centering{StatementExpression, TernaryExpression, LambdaExpression}  & Independent language entities with unique definitions \\
\midrule
\centering{Invocations} & \centering{SuperConstructorInvocation, MethodInvocation, SuperMethodInvocation, SuperMemberReference, ExplicitConstructorInvocation, ArraySelector, AnnotationMethod, MethodReference} & Defines the invocation of a program from another program \\
\midrule
\centering{Declarations} & \centering{TypeDeclaration, FieldDeclaration, MethodDeclaration, ConstructorDeclaration, PackageDeclaration, ClassDeclaration, EnumDeclaration, InterfaceDeclaration, AnnotationDeclaration, ConstantDeclaration, VariableDeclaration, LocalVariableDeclaration, EnumConstantDeclaration, VariableDeclarator} & Declares the existence of an entity in memory and assigns a value to that entity  \\ 
\bottomrule
\end{tabular}
}
\label{table:cstyle}
\end{table}

\subsubsection{Code Stylometry Feature Representation}
\label{subsubsec:feature_rep}
An indispensable component of building predictive models is the transformation of text-based observations into numerical representations. Bengio \textit{et al.}~\cite{bengio2013representation} chronicled the importance of representation learning in predictive modeling. Thus, post-feature selection, we transform selected features to real-valued vectors using Algorithms~1 (statistic) and~2 (neural). 

\textbf{\texttt{Statistical Representation (SR).}} Researchers have demonstrated the utility of applying domain knowledge to retrieve pertinent information by assigning weight values to tokens in a corpus. We leverage this existing knowledge using a feature extractor equation (FE) introduced by Biringa and Kul~\cite{biringa2021automated}. FE numerically transforms selected cstyle features. 

In Algorithm~1, we initialize an empty list ($\mathbb{R}$) to store real-valued vector representations of extracted features and a non-empty list of feature nodes \texttt{(FN)} denoting lexical or syntactic features.  Next, we parse the source files using \texttt{javalang} (a Java programming language parser)~\footnote{\texttt{https://github.com/c2nes/javalang}} and generate $AST$. We traversed the trees AST $t_{i}$ $\in$ $AST$ using a preorder traversal algorithm and selected the features. Following that, we make a subroutine call to FE transformation equation (\texttt{FE}) to convert selected features to vectors: (i) {\tiny$\sum\limits_{i=0}^{n}$} is the count of selected features $(a)$, from $i$ to $n$, and (ii) $\vert x \vert$ is the size of code files in characters. Note, that the inputs for FE are the total count of selected features in a source code and the character length of the aforementioned source code. 

\begin{algorithm}
\SetKwInOut{Input}{\texttt{Input}~}
\SetKwInOut{Output}{\texttt{Output}~}
\Input{\texttt{Repository $\in$ code $\texttt{S} = \{s_{1},s_{2}, \dots,s_{n}$\}} \hfill $\blacktriangleright$ \texttt{Repository source code}}

\Output{${\texttt{v}} =\langle v_{1},v_{2},\dots ,v_{n-1}, v_{n}\rangle$}

$\mathbb{R} \gets \varnothing$ \hfill \texttt{$\blacktriangleright$ Variable to store vectors} \\

\texttt{FE = $-log_{10}(\frac{\sum_{i=0}^{n}(a_i)} {\vert x \vert})$ 
\hfill $\blacktriangleright$ Transformation equation} \\

\texttt{FN} $\gets$ $\{sv_{i},lv_{i},sv_{2},lv_{2},\dots,n$\} \hfill $\blacktriangleright$ \texttt{Selected CStyle features }\\

\textbf{\# \texttt{Selection}} \\
\texttt{\For{$s_{i}$ $\in$ \texttt{S}} {
    $\texttt{curr}$ $\leftarrow$ $\varnothing$\\
    AST  $\gets$  Parse $s_{i}$ $\land$ gen $t_{i}$ $\in$ $AST$ \hfill $\blacktriangleright$ Generate abstract syntax tree (AST) \\
    $a$ $\leftarrow$ Trv $AST_{i}$ $\land$ $\exists$ $\forall$ $n_{i}$ $\in$ FN \hfill $\blacktriangleright$ Traverse AST and select CStyle features \\ \textbf{\# Representation} \\
    $v$ $\leftarrow$ Call FE($a_{i}$, $\vert\mathnormal{\texttt{S}_{i}}\vert$) 
    \hfill $\blacktriangleright$ Transform selected feature using FE \\
    Add $v_{i}$ $\to$ $\texttt{curr}$ \\
    \If {\texttt{syntactic}} { 
        Add $\texttt{curr}$ $\to$ $\mathbb{R}[sv_{i}]$ \hfill \texttt{$\blacktriangleright$ Syntactic feature match}\\ 
    }
    \Else(lexical) {
    Add $\texttt{curr}$ $\to$ $\mathbb{R}[lv_{\texttt{i}}]$ \hfill $\blacktriangleright$ Lexical feature match \\ 
    }
}}
\texttt{\Return $\mathbb{R}$}
\caption{\textbf{\texttt{Statistical Representation of CStyle Features (SR)}}}
\label{algo:stat}
\end{algorithm}

\begin{algorithm}[ht]
\SetKwInOut{Input}{\texttt{Input}~}
\SetKwInOut{Output}{\texttt{Output}~}
\Input{\texttt{Repository $\in$ code $\texttt{S} = \{s_{1},s_{2}, \dots,s_{n}$\}}}
\Output{${\texttt{v}} =\langle v_{1},v_{2},\dots ,v_{n-1}, v_{n}\rangle$}

\texttt{FN} $\gets$ $\{sv_{i},lv_{i},sv_{2},lv_{2},\dots,n$\} \hfill $\blacktriangleright$ \texttt{Selected cstyle features }\\

$\mathbb{R}, \texttt{T},  \texttt{U} \gets \varnothing, \varnothing,  \varnothing $ \hfill \texttt{$\blacktriangleright$ Vector, AST and uncompiled code storages} \\
\textbf{\# \texttt{Selection}} \\

\texttt{\For{\texttt{idx}, \texttt{$s_{i}$} $\in$ \texttt{S}} {
    \If{$s_{i}$ is \texttt{parsable}} {
        $t_{i}$  $\gets$  parse $s_{i}$ $\land$ gen $t_{i}$ \hfill \texttt{$\blacktriangleright$ Generate abstract syntax tree (AST)}\\ 
        Add $t_{i}$  $\to$ \texttt{T}  \hfill \texttt{$\blacktriangleright$ Store AST}\\
    }
    \Else {
    \texttt{U}[idx] $\gets$ $s_{i}$ \hfill \texttt{$\blacktriangleright$ Store uncompiled code}\\
    }
}}

$\texttt{a} \gets \varnothing$ \hfill $\blacktriangleright$ \texttt{Store matched cstyle features} \\ 

$\texttt{SEQL} \gets \texttt{64}$ \hfill $\blacktriangleright$ \texttt{Set maximum sequence length} \\ 

\texttt{\If{\textbf{not} \texttt{U}} {  
    \texttt{\For{$t_{i}$ $\in$ \texttt{T}} {
    $a$ $\leftarrow$ Trv. $t_{i}$ $\land$ $\exists$  $\forall$  $n_{i}$ $\in$ FN \hfill $\blacktriangleright$ \texttt{CStyle feature match} \\
    }} 
}}
\Else{
\textbf{raise} \textit{error} for \texttt{U} \\
}
\textbf{\# \texttt{Representation}} \\
Feed $a$ $:=$ \texttt{Dimensional} $v_{i}$ $\in$ \texttt{32} $\to$ $\mathcal{M}$ \hfill \texttt{$\blacktriangleright$ Train embedding model using features} \\

\texttt{Truncate} $v_{i}$ $\forall$ \texttt{v} $>$ $\land$ \texttt{ZeroPad} $<$ \texttt{SEQL} \hfill \texttt{$\blacktriangleright$ Truncate and pad feature embeddings} \\

\Return $\mathbb{R}$ 
\caption{\textbf{\texttt{Neural Representation of CStyle Features (NR)}}}
\label{algo:cstyleembed}
\end{algorithm}

\textbf{\texttt{Neural Representation (NR).}} We use the word2vec embedding model~\cite{mikolov2013efficient} to represent selected features into real-valued vectors representing neural features of high-dimensional vectors in low-dimensional vector space. We employed the gensim library~\footnote{\texttt{https://radimrehurek.com/gensim/models/word2vec.html}}  to facilitate the generation of vectors and used the default for all parameters except the sequence length and vector dimensions. In algorithm~\ref{algo:cstyleembed}, we transform selected features into embedding vectors. Concisely, selected cstyle features in \texttt{(FN)} are tokenized --- text splitting into independent blocks of characters~\cite{mikolov2013efficient}. Next, we set the maximum sequence (\texttt{SEQL}) length to \texttt{64}, observations $\lessgtr$ \texttt{SEQL} are zero-padded and truncated. Currently, there is no common consensus in the literature regarding what constitutes an appropriate embedding size, only predominantly task-dependent suggestions~\cite{biringa2022short}. However, after experimenting with embeddings of different sizes, we found that a \texttt{64} sequence length suffices in maintaining a permissible relationship between a model's predictive prowess and training time in this case. Finally, we return generated feature embeddings. 

\subsection{Phase 3: Functional Tests' Times Extraction}
\label{subsec:automated_testing}
Recall in our problem definition (see Section~\ref{sec:motivating_bg}) we executed functional test cases, after which we logged the test times to derive the performance of programs, ensuring correctness. The test times are microbenchmarks (dependent variable) fed to predictive models. We utilized Travis CI (\texttt{TCI})~\cite{citools} and Maven Surefire (\texttt{MVNS})~\cite{rinian2013introduction} to extract the execution test times. \texttt{TCI} is an open-source continuous integration service used to build and test software hosted on GitHub and other source code management and distributed version control platforms. We created a \texttt{.travis.yml} file --- compiles and executes functional test scripts present in candidate repositories, such as unit tests --- to link the candidate repository and testing components, after which we deployed and ran repository test scripts. Following that, we executed test cases and logged the test time. Maven Surefire (\texttt{MVNS}) is the most frequently used automation tool for building and testing Java software. We retrieved the  \texttt{MVNS} testing tool from the Apache website and performed the same linking operation used for the \texttt{TCI} framework.

\subsection{Phase 4: Continuous Performance Prediction}
\label{subsec:models}
Continuously predicting the performance impact of syntactically correct source code to software is a core feature of \texttt{PACE}. Directly checking the ground truth performance of isolated or local code updates using testing and profiling tools is possible. However, it is a deviation from \texttt{PACE's} objective, which entails estimating what updates and potentially low-performing code contribute to the overall software performance before being pushed to the primary or remote repository. 

\begin{figure}[ht]
    \centering
    \captionsetup{justification=centering}
    \subfloat[Supervised Learning]
    {{\includegraphics[width=4.5cm]{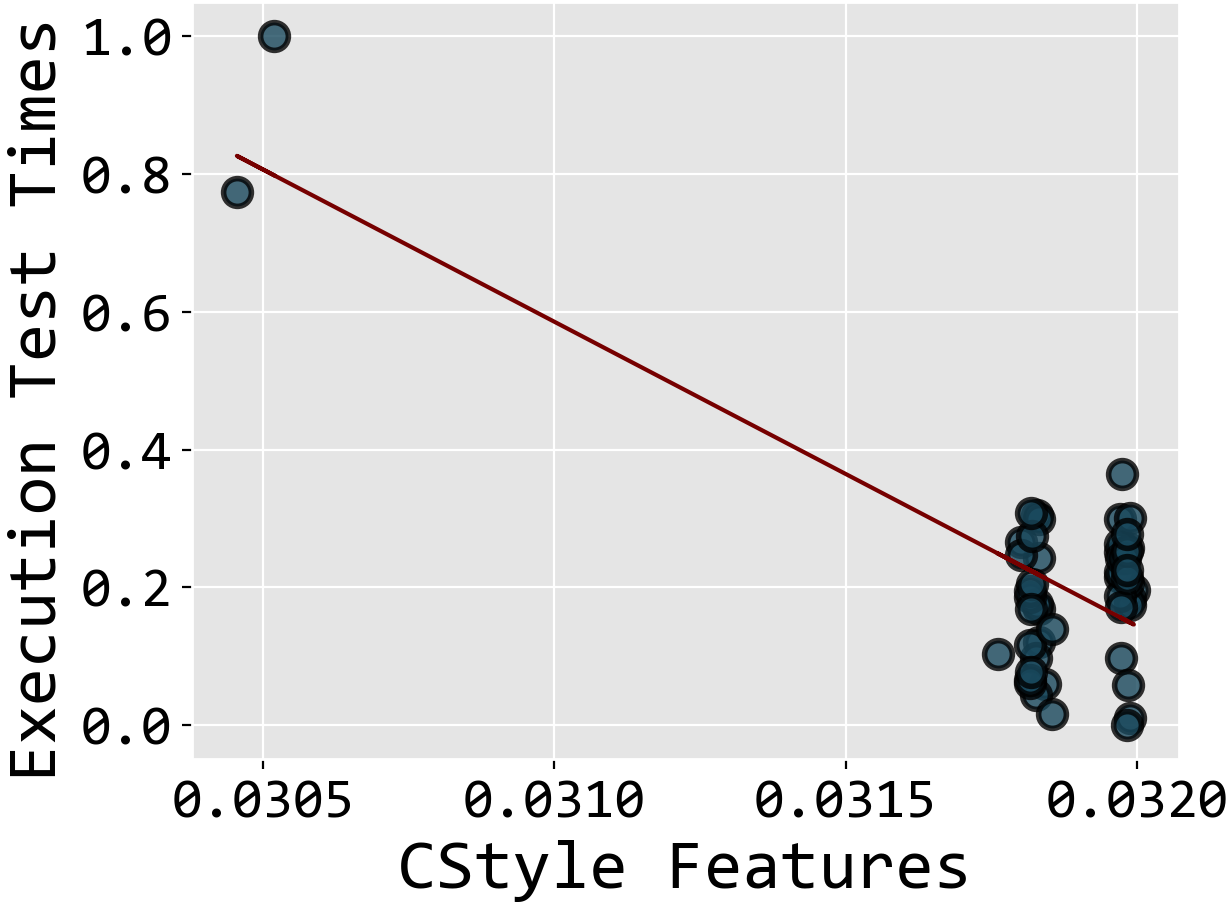}}}
    \subfloat[Multiple Regression]
    {{\includegraphics[width=4.5cm]{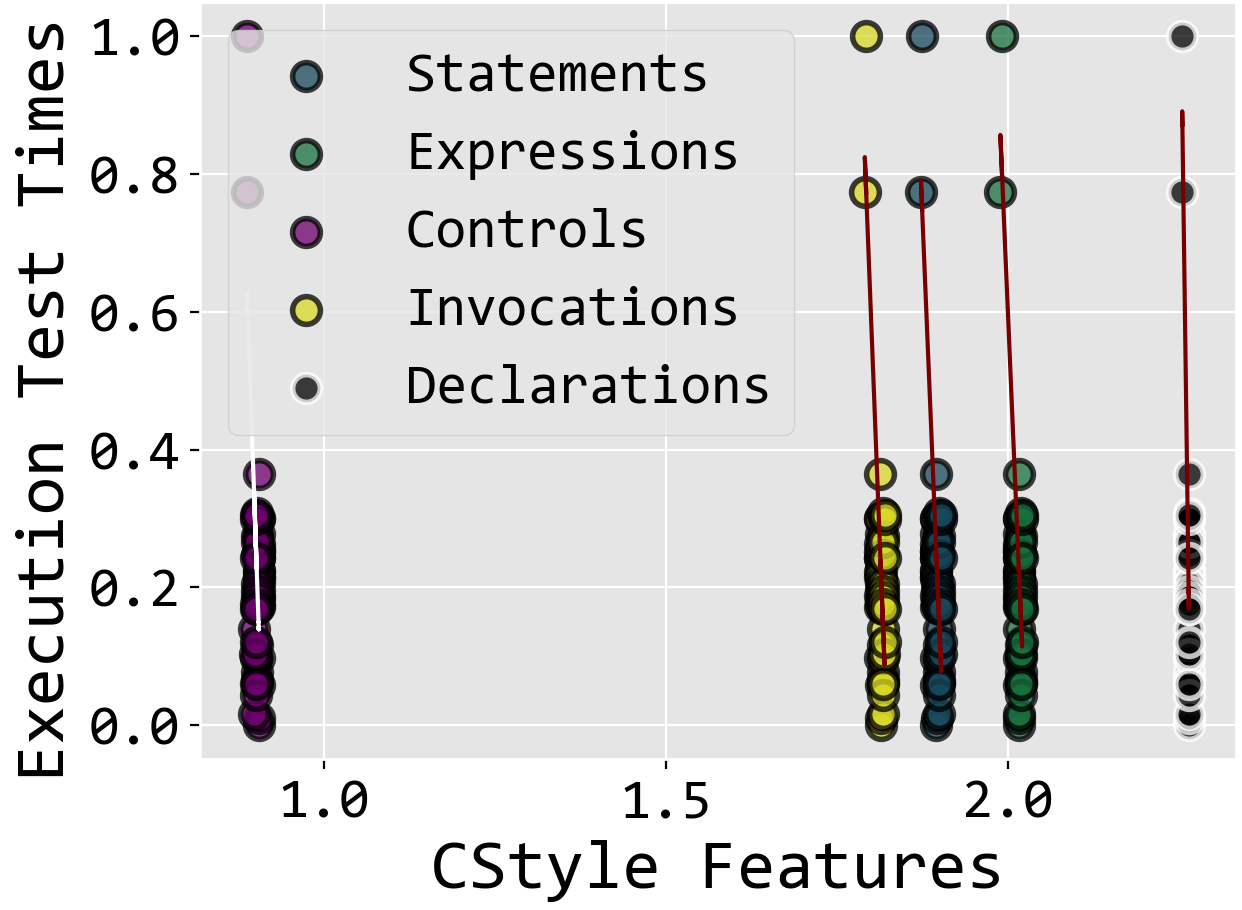}}}
    \subfloat[Univariate Multiple Regression]
    {{\includegraphics[width=4.5cm]{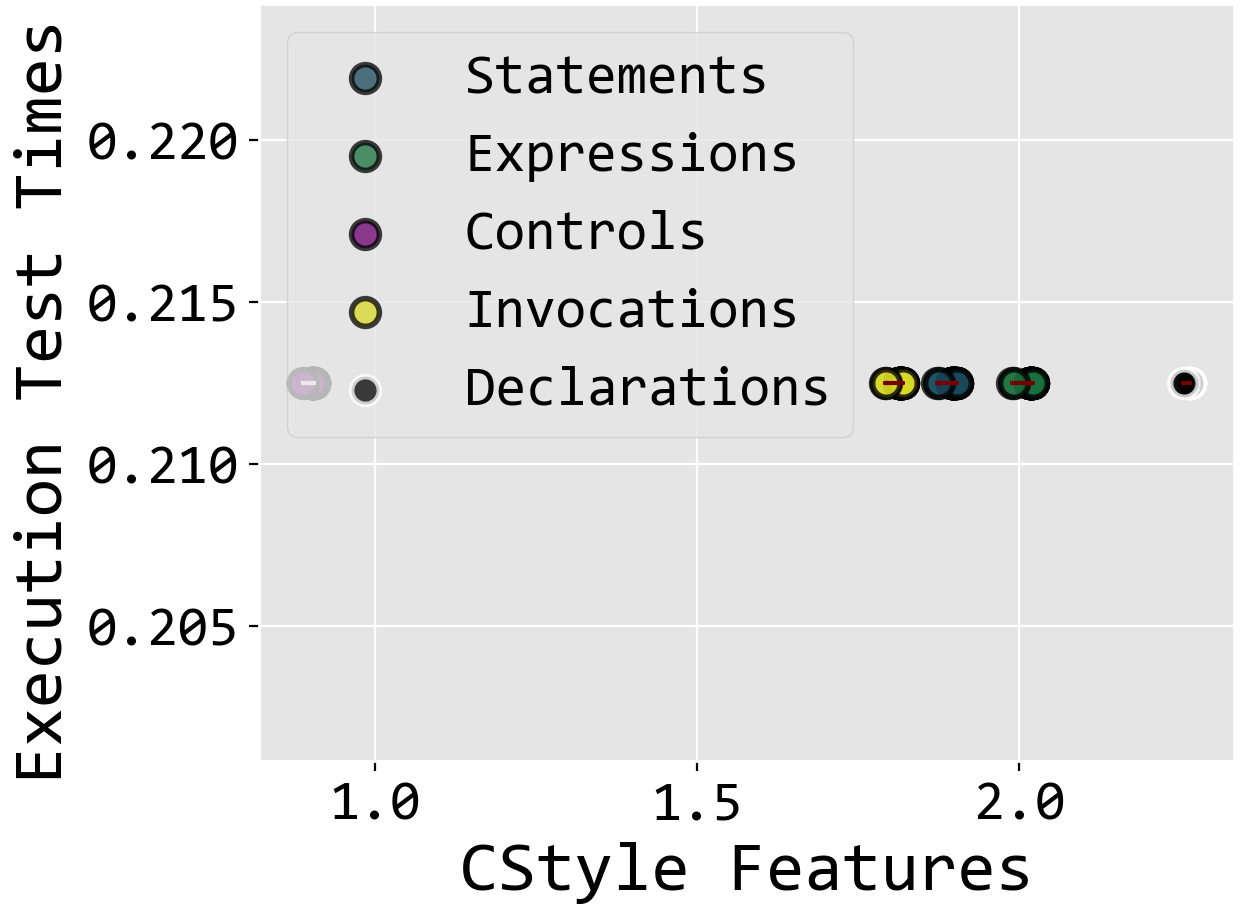}}}
    \vspace{-2mm}
    \caption{Problem Formulation for Regression Model Selection}
    \label{fig:pf}
\end{figure}

To achieve \texttt{PACE's} objective, we employed a collection of regression models~\cite{doran2017does}, which are supervised machine learning algorithms used to model the relationship between single or multiple independents (predictor) and a dependent variable (response). Regression models are used to predict continuous data values, which is fitting for predicting software execution time performance in this case. We design our model by attentively framing the problem, given that: (i) it is a supervised learning problem, given that our training samples are labeled, for example, mapping the execution test times target to cstyle features (see Figure~\ref{fig:pf}a for a visual depiction), (ii) it is a multiple regression problem, given that we utilize several feature classes to perform predictions (see Figure~\ref{fig:pf}b), and (iii) it is a univariate regression problem, given that we are attempting to predict a single execution time performance for each commit (see Figure~\ref{fig:pf}c). Following cstyle feature selection and representation, we trained five univariate multi-regression models.

\begin{table}[ht]
\caption{A Taxonomy of Regression Models for Continuous Performance Prediction}
\resizebox{0.99\linewidth}{!}{
\begin{tabular}
{{ p{0.12\linewidth} | p{0.20\linewidth} | p{0.30\linewidth} | p{0.50\linewidth} }}
\toprule
\centering{\textbf{Learning}} & \centering{\textbf{Class}} & \centering{\textbf{Model}} & \textbf{Short Description} \\
\toprule
\centering{Deep} & \centering{Neural Network} & \centering{Multilayer Perception (MLP)} & A feed-forward artificial neural network \\
\midrule
\centering{Error} & \centering{Support Vector Machines} & \centering{Support Vector Regressor (SVR)} & A non-probabilistic model that efficiently utilizes a decision boundary to maximize discrete value margins \\
\midrule
\centering{Information} & \centering{Decision Trees} & \centering{Random Forest Regressor (RFR)} & A collection of decision trees built during training referred to as estimators \\ 
\midrule
\centering{Probabilistic} & \centering{Bayesian Network} & \centering{Bayesian Ridge (BR)} & A Bayesian probabilistic model derived from Bayes theorem \\
\midrule
\centering{Lazy} & \centering{Distance} & \centering{k-Nearest Neighbors Regressor (kNN)} & A lazy model that employs distance between data points in deriving insights \\
\bottomrule
\end{tabular}
}
\label{table:reg} 
\end{table}

\section{Evaluation}
\label{sec:evaluation}
This section describes selected evaluation metrics (see Section~\ref{subsec:eval_metrics}) and case study datasets (see Section~\ref{subsec:datasets}).

\noindent \textbf{\texttt{Setup.}}
We conducted experiments using an Apple commodity laptop with an Apple M1 8-core CPU, 16-core Neural Engine, 8 GB RAM, and macOS Big Sur (version 11).

\noindent \textbf{\texttt{Reproducibility.}}
We have made all source code, results, and datasets publicly available~\footnote{\texttt{https://github.com/PADLab/PACE}} with command line arguments to reproduce the results presented in this paper.

\subsection{Metrics}
\label{subsec:eval_metrics}
We comprehensively evaluate our models' performance using classical ML regression evaluation metrics~\cite{geron2019hands, pandey2020seir}. We employ Root Mean Squared Error \texttt{(RMSE)}, Mean Squared Error \texttt{(MSE)}, Mean Absolute Error \texttt{(MAE)}, and Root Mean Squared Logarithmic Error to assess the predictive performance of models.

\begin{itemize}
    \item \textbf{\texttt{RMSE}}: Adopts a quadratic approach that measures the squared root of the mean squared difference between predicted and actual observations.
     \item \textbf{\texttt{MSE}}: The average squared distance between the actual observations and predicted.
     \item \textbf{\texttt{MAE}}: Calculates the mean error sizes in predictions and does not consider the directional relationship between predicted and actual. It is only concerned with the absolute difference.
     \item \textbf{\texttt{RMSLE}}: Calculates the log difference between actual and predicted values.
\end{itemize}

\noindent In selecting regression metrics, we are motivated by three applications.
\begin{itemize}
\item \textbf{\texttt{Application 1:}} For a conservative use case (underestimating code performance predictions (CPP)), we recommend applying RMSE or MAE.
\item \textbf{\texttt{Application 2:}} For a liberal use case (overestimating (CPP)), we recommend applying MSE.
\item \textbf{\texttt{Application 3:}} For a centrist use case, we recommend applying RMSLE.
\end{itemize}

{\small
\begin{align*}
\texttt{\textbf{RMSE}}=\sqrt{(\frac{1}{n})\sum_{i=1}^{n}(y_{i} - x_{i})^{2}}
\;
\texttt{\textbf{MSE}}=\sum_{i=1}^{D}(x_i-y_i)^2 \;
\texttt{\textbf{MAE}}=(\frac{1}{n})\sum_{i=1}^{n}\left | y_{i} - x_{i} \right | \;
\texttt{\textbf{RMSLE}} = \sqrt{\frac{1}{N}\sum_{i=1}^{n}\left ( \log{yi} - \log{\hat{yi}} \right )^2}
\end{align*}
}

\subsection{Datasets}
\label{subsec:datasets}
\noindent \textbf{\texttt{AddressBook Dataset (ABD).}} We obtain the AddressBook dataset from the Addressbook~\footnote{\texttt{https://github.com/se-edu/addressbook-level2}} repository - a command-line interface program used to teach software engineering students object-oriented programming. We applied \texttt{RPiTS} to extract five commits. We preprocess the repository by only selecting Java files, with commit branches that comprise 210 total, 42 $\mu$ Java files, and 10,870 lines of code. We concatenate the source code to obtain observations. Finally, we join git branches to the execution time microbenchmarks from \texttt{TCI} to create the final dataset.

\noindent \textbf{\texttt{DSpace Dataset (DSD).}}
We extracted the DSpace dataset from DSpace~\footnote{\texttt{https://github.com/DSpace/DSpace}} repository - an open-source web application software for providing digital resources to thousands of organizations and institutions worldwide and has a large and active developer community. The dataset extracted from the original repository using \texttt{RPiTS} contains 50 commits, 127,157 total files, and 2543 $\mu$ Java files, 32,198,822,697 lines of code. We create the final dataset by joining git branches to execution time microbenchmarks derived from \texttt{MVNS}.

\section{Experiments}
\label{sec:result}
\begin{table}[!htbp]
\caption{Predictive models for continuous performance prediction using statistical and neural representations from code-stylometry features. Commit ($c_{n}$) denotes repository CCS, where \{$c_{n}, c_{n-1}\} \in$ \{training, testing\} sets. \texttt{RMSE, MSE, MAE, and RMSLE} denotes performance metrics used to evaluate the predictive power of implemented models as discussed in Section~\ref{subsec:eval_metrics}. \texttt{AVG} denotes predictor performance across metrics. \texttt{\#SC} denotes the total source files for each CCS}
\resizebox{0.92\linewidth}{!}{
\begin{tabular}
{{p{0.11\linewidth} | p{0.11\linewidth} | p{0.06\linewidth} p{0.06\linewidth} p{0.07\linewidth} p{0.07\linewidth} p{0.06\linewidth} | p{0.07\linewidth} p{0.06\linewidth} p{0.07\linewidth} p{0.07\linewidth} p{0.06\linewidth}
}}
\toprule
  \multicolumn{12}{c}{\textbf{\texttt{\texttt{ABD}}}} \\
  \toprule
\multicolumn{5}{r}{\textbf{\texttt{SR Features}}} 
&&
\multicolumn{6}{c}{\textbf{\texttt{NR Features}}} \\
\toprule
 \textbf{\texttt{Predictor}} & \textbf{\texttt{Commit}} & \textbf{\texttt{RMSE}} & \textbf{\texttt{MSE}} & \textbf{\texttt{MAE}} & \textbf{\texttt{RMSLE}} & \textbf{\texttt{AVG}} & \textbf{\texttt{RMSE}} & \textbf{\texttt{MSE}} &\textbf{\texttt{\texttt{MAE}}} & \textbf{\texttt{RMSLE}} & \textbf{\texttt{AVG}}
\\  
\toprule
\texttt{PACE-MLP} & \text{$c_{n}, c_{n-1}$} & 0.5280 &  0.2788 & 0.4923 & 0.1674  & 0.3666 & 0.2391 & 0.0571 & 0.2284 & 0.0783  & 0.1507 \\  
\midrule
 & \text{$c_{n-1}, c_{n-2}$} & 0.2088 & 0.0436 & 0.1903 & 0.0678  & 0.1276 & 0.2234 & 0.0499 & 0.2004 & 0.0735  & 0.1368 \\
\midrule
 & \text{$c_{n-2}, c_{n-3}$} & 0.0980 & 0.0096 & 0.0758 & 0.0309  & 0.0536 & 0.0539 & 0.0029 & 0.0267 & 0.0174  & 0.0252 \\
\midrule
 & \text{$c_{n-3}, c_{n-4}$} & 0.1220 & 0.0148 & 0.0968 & 0.0374  & 0.0678 & 0.0867 & 0.0075 & 0.0771 & 0.0275  & 0.0497 \\
 \midrule
  & & & & & \textbf{\texttt{AVG}} $\blacktriangleright$  & 0.1539 & & & & \textbf{\texttt{AVG}} $\blacktriangleright$ & 0.0906 \\
 \midrule
\texttt{PACE-SVR} & \text{$c_{n}, c_{n-1}$} & 0.2199 & 0.0483 & 0.2199 & 0.0698  & 0.1395 & 0.2199 & 0.0483 & 0.2199 & 0.0698  & 0.1395 \\   
\midrule
& \text{$c_{n-1}, c_{n-2}$} & 0.1800 & 0.0324 & 0.1800 & 0.0575  & 0.1124 & 0.1800 & 0.0324 & 0.1800 & 0.0575  & 0.1124 \\  
\midrule
& \text{$c_{n-2}, c_{n-3}$} & 0.0200 & 0.0004 & 0.0200 & 0.0061  & 0.0116 &  0.0200 & 0.0004 & 0.0200 & 0.0061  & 0.0116 \\   
\midrule
& \text{$c_{n-3}, c_{n-4}$} & 0.0700 & 0.0049 & 0.0700 & 0.0218  & 0.0416 & 0.0700 & 0.0049 & 0.0700 & 0.0218  & 0.0416 \\   
\midrule
  & & & & & \textbf{\texttt{AVG}} $\blacktriangleright$ & 0.0763 & & & & \textbf{\texttt{AVG}} $\blacktriangleright$ & 0.0763 \\
\hline
\rowcolor{myblue}
\texttt{PACE-kNN} & \text{$c_{n}, c_{n-1}$} & 0.2199 & 0.0483 & 0.2199 & 0.0698 & 0.1395 & 0.2199 & 0.0483 & 0.2199 & 0.0698  & 0.1395 \\  
\hline
\rowcolor{myblue}
& \text{$c_{n-1}, c_{n-2}$} & 0.1800 & 0.0324 & 0.1800 & 0.0575  & 0.1124 & 0.1800 & 0.0324 & 0.1800 & 0.0575  & 0.1124 \\  
\hline
\rowcolor{myblue}
& \text{$c_{n-2}, c_{n-3}$} &  0.0200 & 0.0004 & 0.0200 & 0.0061  & 0.0116 & 0.0200 & 0.0004 & 0.0200 & 0.0061  & 0.0116 \\  
\rowcolor{myblue}
\hline
& \text{$c_{n-3}, c_{n-4}$} & 0.0700 & 0.0049 & 0.0700 & 0.0218  & 0.0416 & 0.0700 & 0.0049 & 0.0700 & 0.0218  & 0.0416 \\  
\hline
\rowcolor{myblue}
  & & & & & \textbf{\texttt{AVG}} $\blacktriangleright$ & 0.0763 & & & & \textbf{\texttt{AVG}} $\blacktriangleright$ & 0.0763 \\
\hline
\texttt{PACE-RFR} & \text{$c_{n}, c_{n-1}$} & 0.2199 & 0.0483 & 0.2199 & 0.0698  & 0.1395 &  0.2199 & 0.0483 & 0.2199 & 0.0698  & 0.1395  \\   
\midrule
& \text{$c_{n-1}, c_{n-2}$} & 0.1800 & 0.0324 & 0.1800 & 0.0575  & 0.1124 & 0.1800 & 0.0324 & 0.1800 & 0.0575  & 0.1124 \\  
\midrule
& \text{$c_{n-2}, c_{n-3}$} & 0.0200 & 0.0004 & 0.0200 & 0.0061  & 0.0116 & 0.0200 & 0.0004 & 0.0200 & 0.0061  & 0.0116 \\  
\midrule
& \text{$c_{n-3}, c_{n-4}$} & 0.0700 & 0.0049 & 0.0700 & 0.0218  & 0.0416 & 0.0700 & 0.0049 & 0.0700 & 0.0218  & 0.0416 \\  
\midrule
& & & & & \textbf{\texttt{AVG}} $\blacktriangleright$ & 0.0763 & & & & \textbf{\texttt{AVG}} $\blacktriangleright$ 
 & 0.0763 \\
\midrule
\texttt{PACE-BR} & \text{$c_{n}, c_{n-1}$} & 0.2200 &  0.0484 & 0.2200 & 0.0698  & 0.1395 & 0.2200 & 0.0484 & 0.2200 & 0.0698  & 0.1395 \\  
\midrule
& \text{$c_{n-1}, c_{n-2}$} & 0.1800 & 0.0324 & 0.1800 & 0.0575  & 0.1124 & 0.1800 & 0.0324 & 0.1800 & 0.0575  & 0.1124 \\  
\midrule
& \text{$c_{n-2}, c_{n-3}$} & 0.0200 & 0.0004 & 0.0200 & 0.0061  & 0.0116 & 0.0200 & 0.0004 & 0.0200 & 0.0061  & 0.0116 \\  
\midrule
& \text{$c_{n-3}, c_{n-4}$} & 0.0699 & 0.0048  & 0.0699 & 0.0218  & 0.0416 & 0.0699 & 0.0048  & 0.0699 & 0.0218  & 0.0416 \\  
\midrule
  & & & & & \textbf{\texttt{AVG}} $\blacktriangleright$  & 0.0763 & & & & \textbf{\texttt{AVG}} $\blacktriangleright$  & 0.0763 \\
 \bottomrule
\end{tabular}
}
\label{table:RQ1} 
\end{table}

 
\subsection{RQ1. What is the predictive accuracy of code performance given an update?}
\begin{figure}[!htbp] 
    \centering
    \captionsetup{justification=centering}
    \subfloat[Performance (MSE, MAE and RMSLE)]
    {{\includegraphics[width=5cm]{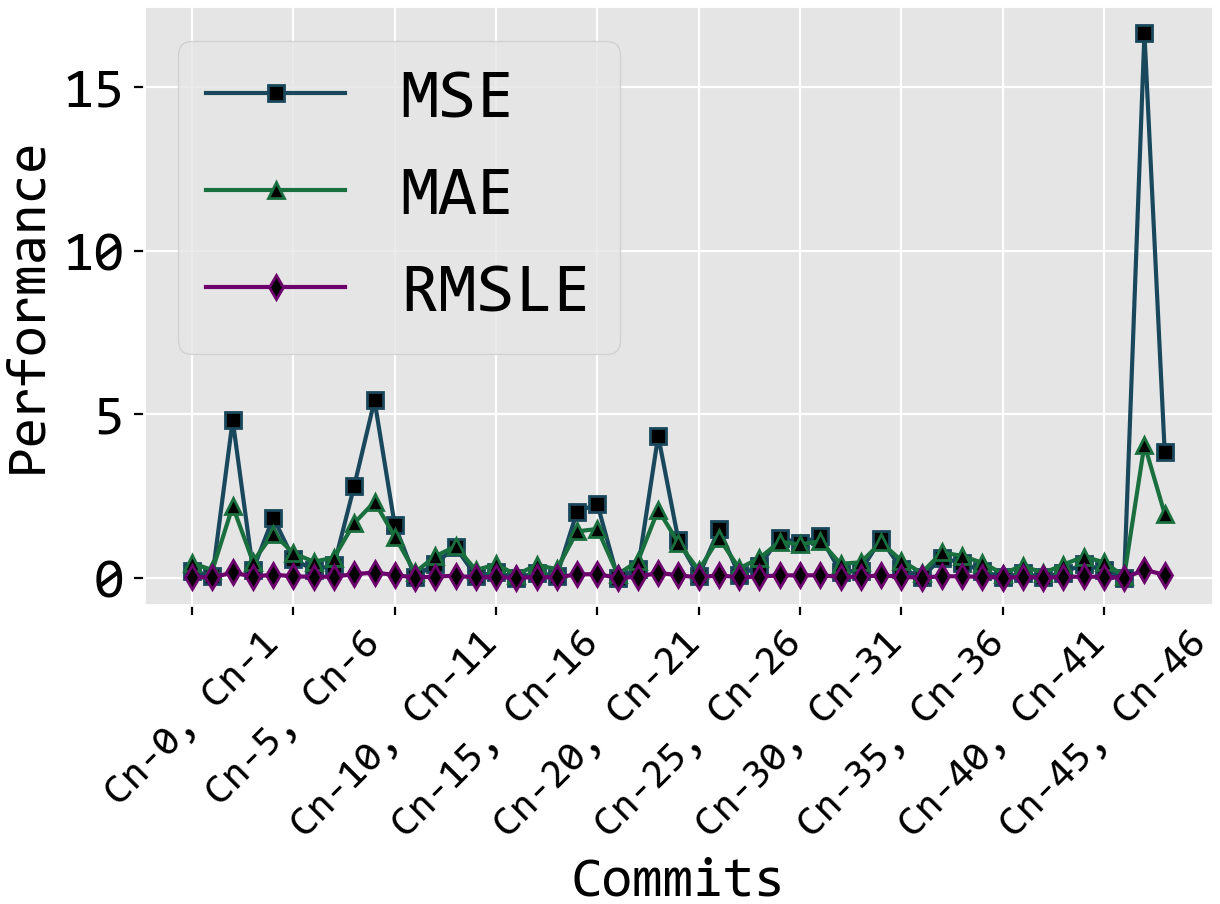}}}
    \;
    \subfloat[Performance (AVG)]
    {{\includegraphics[width=5cm]{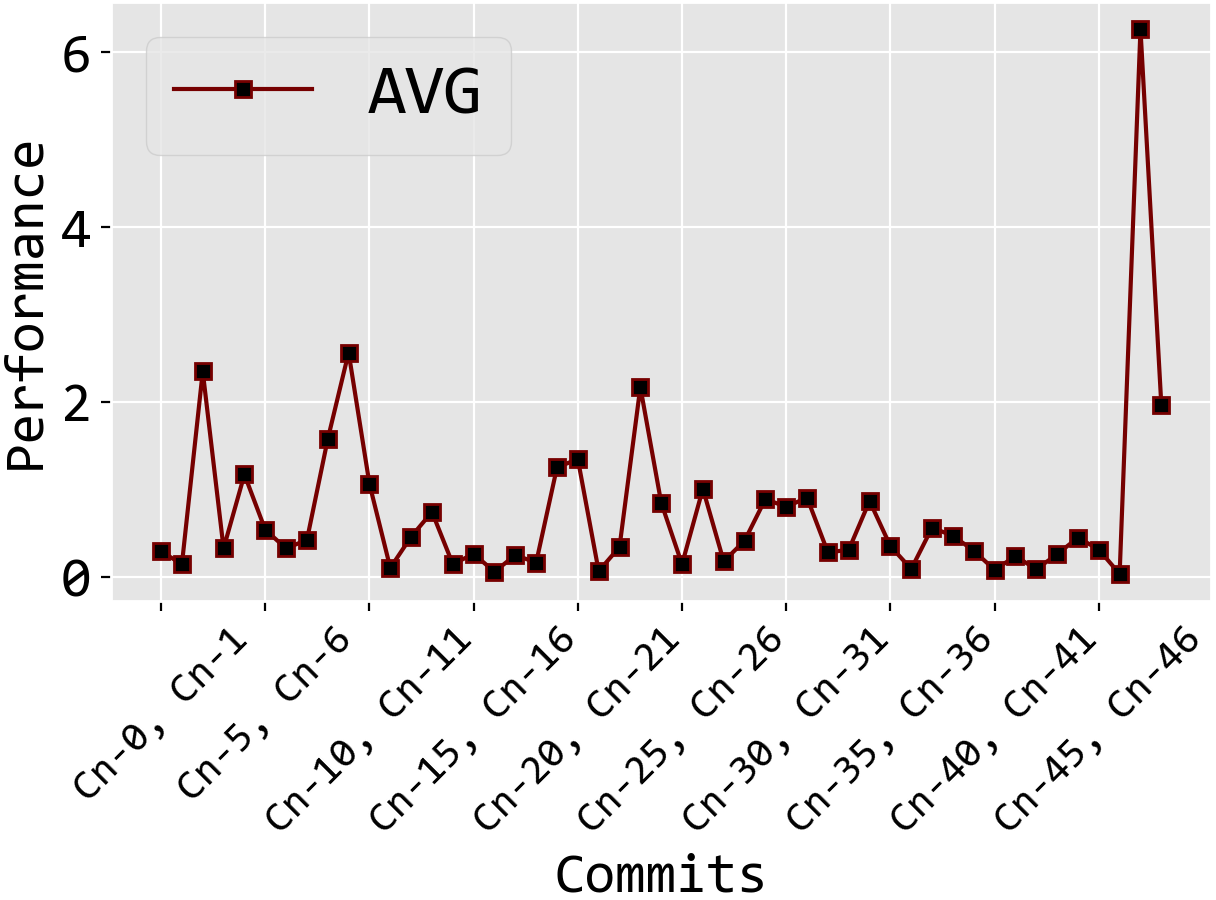}}}
    \vspace{-2mm}
    \caption{PACE-kNN for continuous performance prediction on DSD. To deduplicate generated results from RMSE in Table~\ref{table:rq1}, we use only MSE, MAE, RMSLE, and the average value of the Features (SR, NR)}
    \label{fig:RQ1_perf}
\end{figure}

We answer this question by first implementing regression models with \texttt{SR} and \texttt{NR} input-feature vectors using the 5-commits \texttt{ABD} in Table~\ref{table:RQ1}, after which we select the best-performing model (\texttt{PACE-kNN}). The selected (best performing) model is then exclusively used as the primary predictor on the 50-commits \texttt{DSD} in Figure~\ref{fig:RQ1_perf}. The predictive ((\texttt{0.0763, 0.7366) $\mu$(RMSE, MSE, MAE, RMSLE (RMMR)) $\in$ SR $\land$ NR}) performances of \texttt{PACE-kNN's} is slightly better than other models when prediction throughput (\texttt{0.0063 $\mu$ seconds (sec) $\in$ SR $\land$ NR}) is factored in. Hence, we select \texttt{PACE-kNN} as our primary predictor. Interestingly, all models except \texttt{PACE-MLP} perform similarly to the fourth decimal place across metrics for both features. 

Thus, outside the model's prediction times, there is no marketable difference in performance between the statistic and neural representation of cstyle features. Furthermore, we observe that the single (\texttt{($c_{n-48}, c_{n-49}$) = 6.2660 $\mu$(RMMR))} outlier predictive performance in the \texttt{DSD} has both the numerically highest target variable (4.081 sec) and source file (46 files) difference, where (\texttt{($c_{n-48} = 13.528, c_{n-49} = 17.609$) sec $\land$ ($c_{n-48} = 2559, c_{n-49} = 2605$)} the total number of files) $\in$ (\texttt{train, test}) sets respectively, indicating that a commit with a disproportionate code introduction or deduction can skew the results of predictors. However, as demonstrated in (\texttt{($c_{n-49}, c_{n-50}$) = 1.9669 $\mu$(RMMR)}), the model's performance improves with a subsequent commit. 

\vspace{-2mm}
\vspace{-2mm}

\begin{figure}[!htbp]
    \centering
    \captionsetup{justification=centering}
    \subfloat[PACE-MLP]
    {{\includegraphics[width=4.5cm]{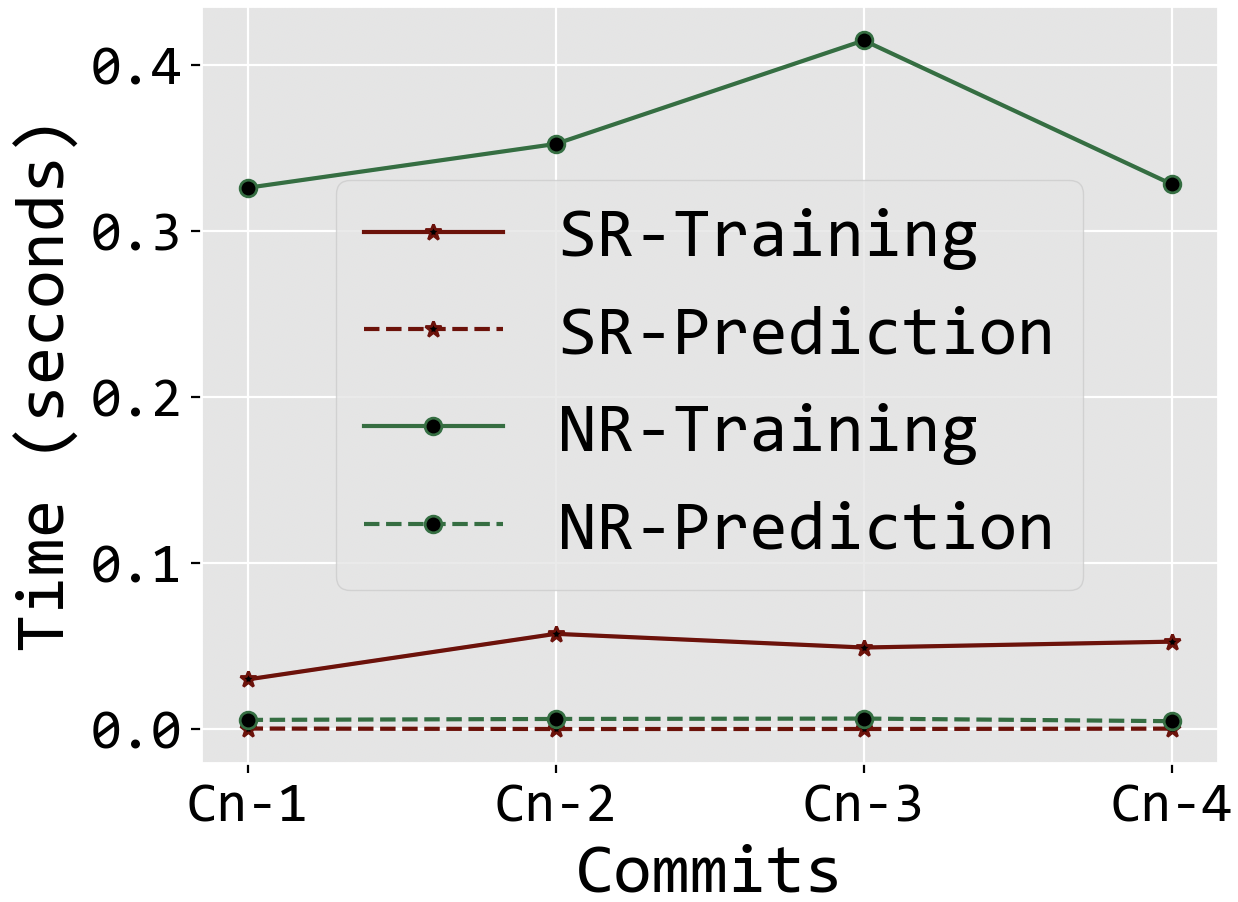}}}
    \subfloat[PACE-SVR]
    {{\includegraphics[width=4.5cm]{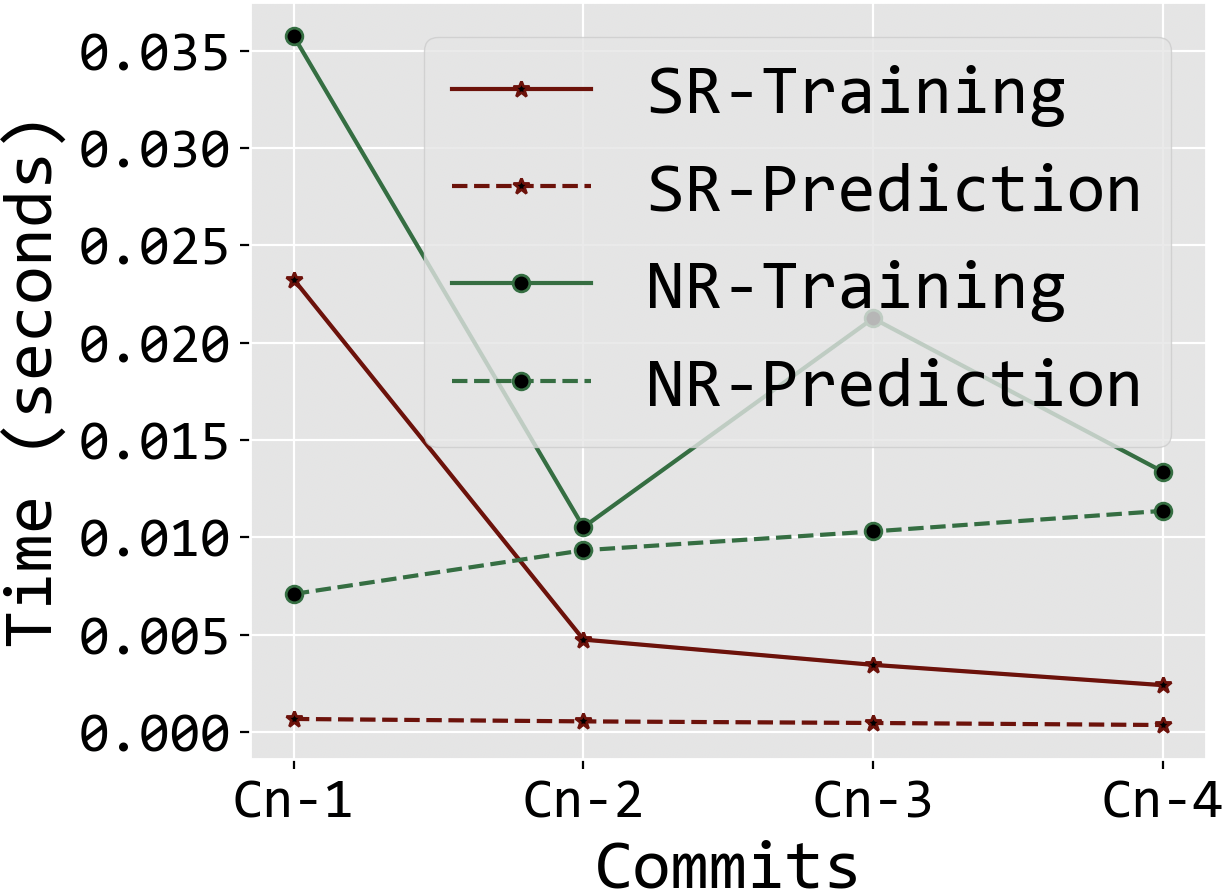}}}
    \subfloat[PACE-kNN(\texttt{ABD})]
    {{\includegraphics[width=4.5cm]{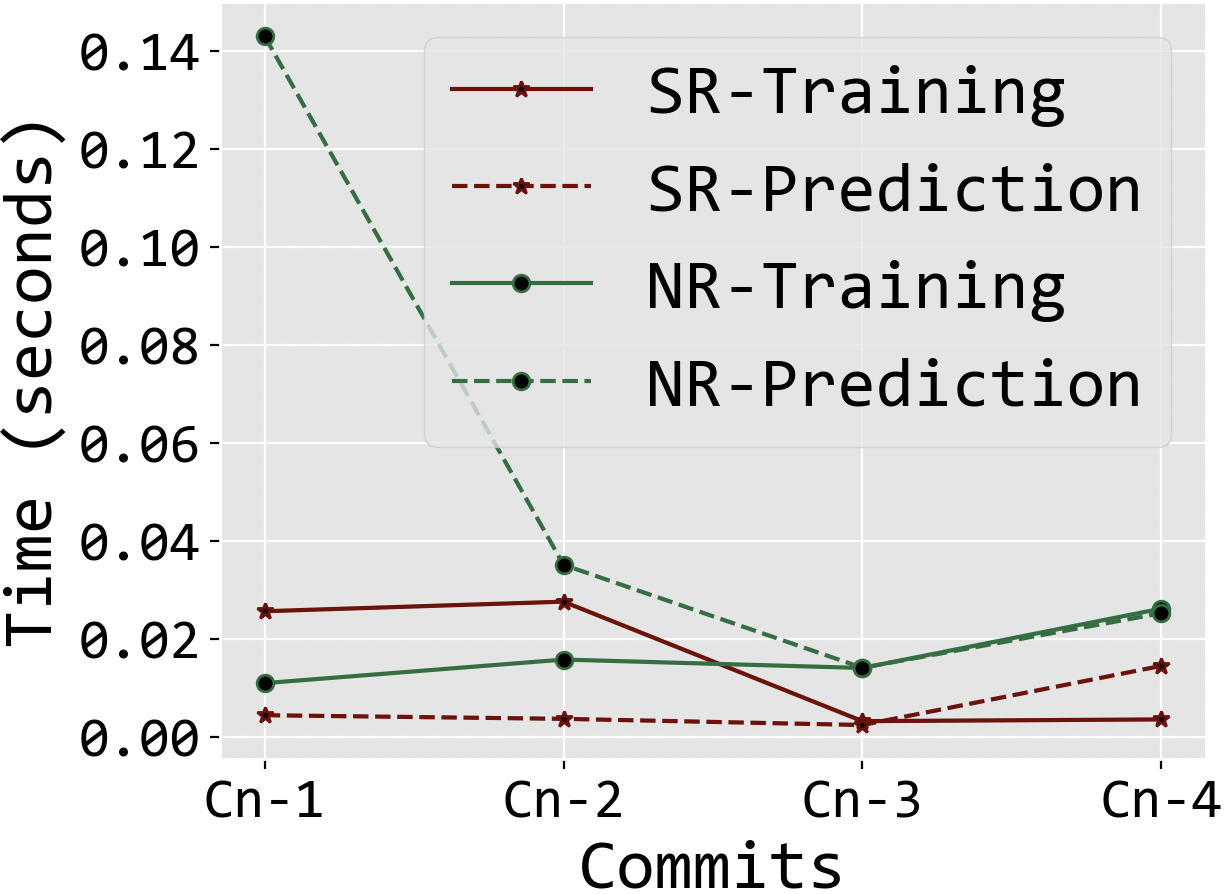}}}
    \vspace{-2mm}
    \caption{ Training (TT) and Prediction (PT) Throughput of Predictors for SR and NR Features on \texttt{ABD}. \\
    PACE-MLP: $\mu$(TT, PT) = 0.0239 sec $\in$ SR. $\mu$(TT, PT) = 0.1807 sec $\in$ NR. $\mu$(SR, NR) = 0.1023 sec \\ 
    PACE-SVR: $\mu$(TT, PT) = 0.0044 sec $\in$ SR. $\mu$(TT, PT) = 0.0148 sec $\in$ NR. $\mu$(SR, NR) = 0.0096 sec \\
    PACE-kNN: $\mu$(TT, PT) = 0.0106 sec $\in$ SR. $\mu$(TT, PT) = 0.0356 sec $\in$ NR. $\mu$(SR, NR) = 0.0231 sec}
    \label{fig:tp_mod}
\end{figure}

\begin{figure}[!htbp]
    \centering
    \captionsetup{justification=centering}
    \subfloat[PACE-RFR]
    {{\includegraphics[width=4.5cm]{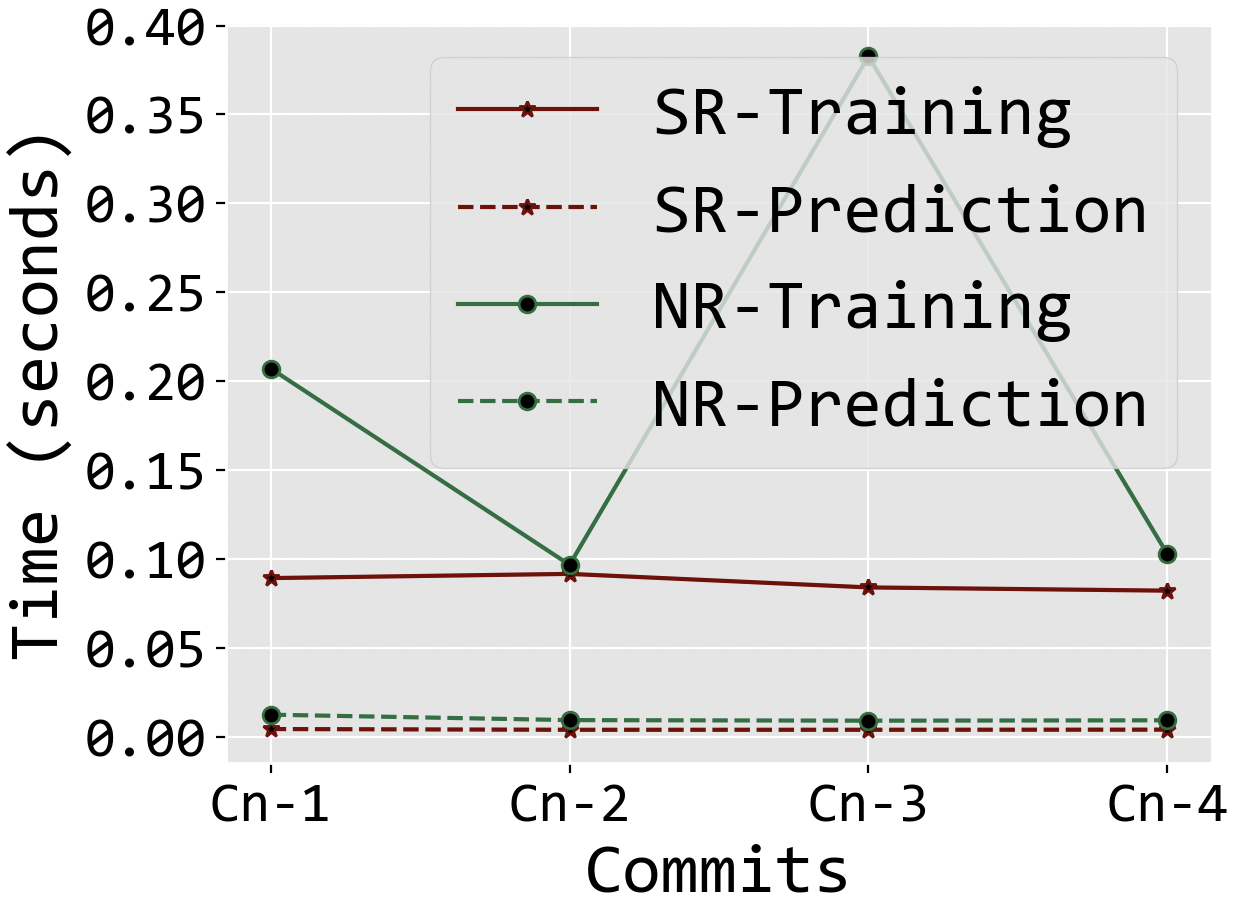}}}
    \subfloat[PACE-BR]
    {{\includegraphics[width=4.5cm]{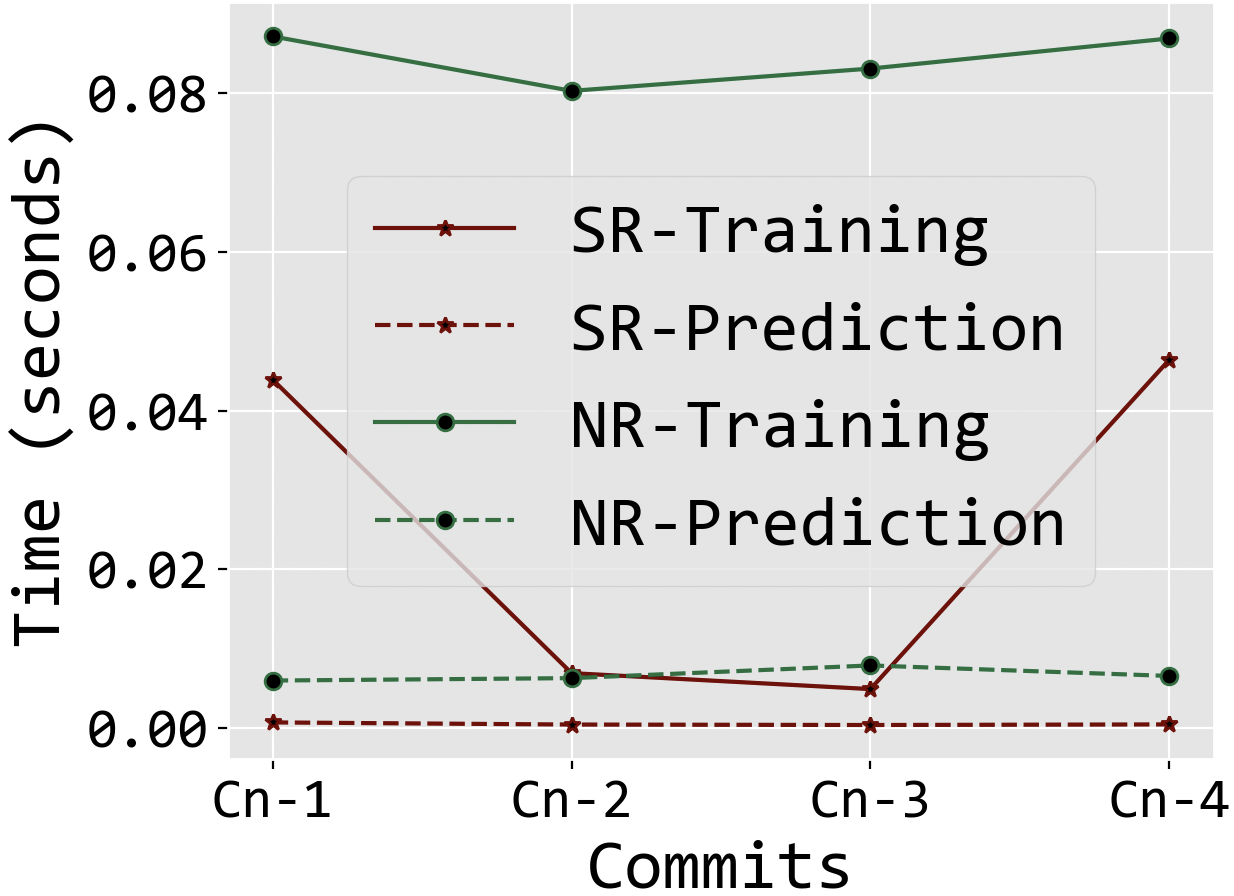}}}
    \subfloat[PACE-kNN(DSD)]
    {{\includegraphics[width=4.5cm]{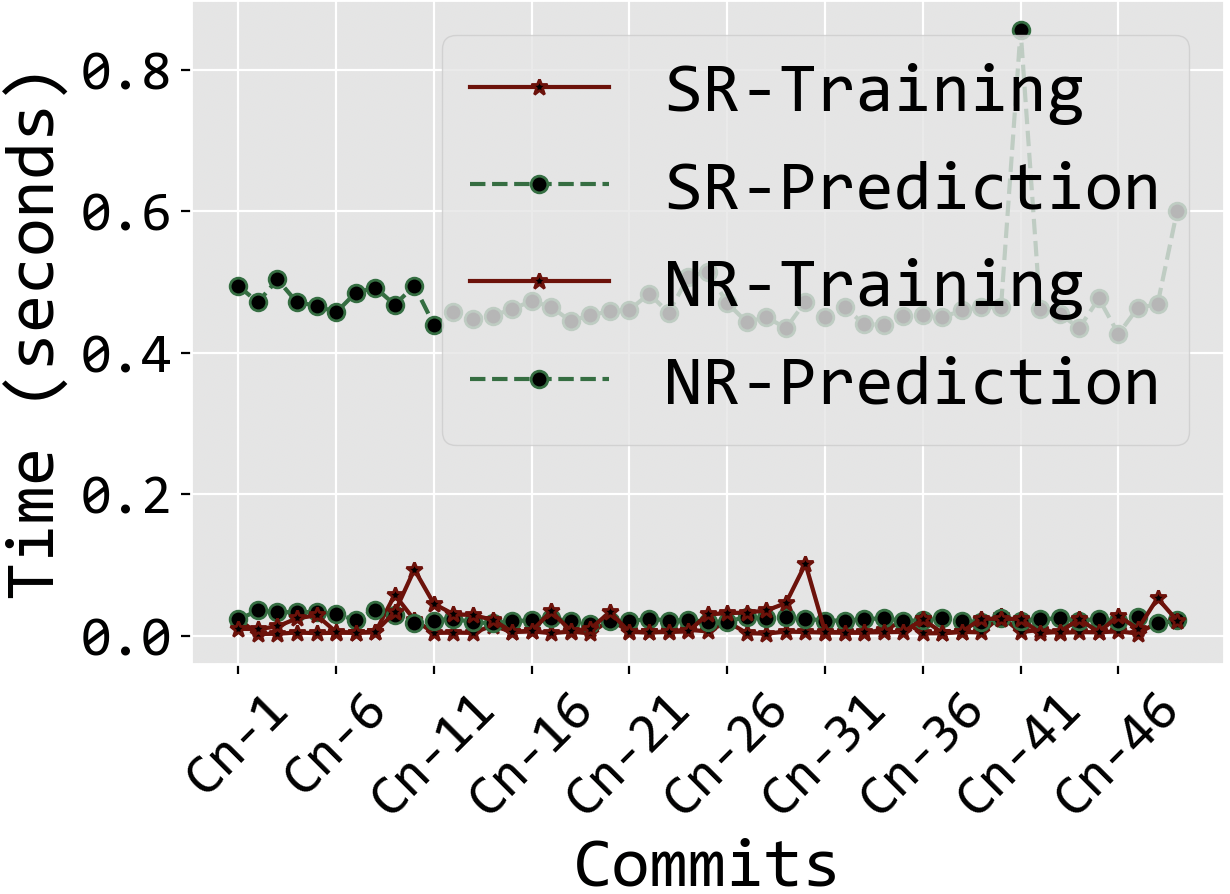}}}
    \vspace{-2mm}
    
    \caption{Training (TT) and Prediction (PT) Throughput of Predictors for SR and NR Features on \texttt{ABD} and \texttt{DSD}. 
    \\
    PACE-RFR: $\mu$(TT, PT) = 0.0454 $\in$ SR. $\mu$(TT, PT) = 0.1037 $\in$ NR. $\mu$(SR, NR) = 0.0745
    \\
    PACE-BR: $\mu$(TT, PT) = 0.0130 $\in$ SR. $\mu$(TT, PT) = 0.0455 $\in$ NR. $\mu$(SR, NR) = 0.0292 \\
    PACE-kNN(SR): $\mu$(TT, PT) = 0.0075. PACE-kNN(NR): $\mu$(TT, PT) = 0.1037. $\mu$(SR, NR) = 0.0745
    }
    \label{fig:tp_mod1}
\end{figure}

Figures~\ref{fig:tp_mod} and~\ref{fig:tp_mod1} visualizes the training and prediction throughput of all models. \texttt{PACE-MLP} has the fastest prediction time with a (\texttt{0.0003}$\mu$ sec). However, with a 0.1222 (\texttt{$\mu$(RMMR)} $\in$ \texttt{SR} $\land$ \texttt{NR}), it underperforms \texttt{PACE-kNN} and other models by 46\%. \texttt{PACE-BR} (\texttt{0.0004}$\mu$ sec) comes a close second, followed by \texttt{PACE-SVR} (\texttt{0.0005}$\mu$ sec)  and \texttt{PACE-RF} (\texttt{0.0005}$\mu$ sec). We also present the predictors' latency (time elapsed to train and test predictors using \texttt{SR} and \texttt{NR} features). \texttt{PACE-SVR ($\mu$(SR, NR) = 0.0096 sec)} has the lowest latency pipeline on \texttt{ABD}, outperforming \texttt{PACE-kNN (0.0231 sec)}, \texttt{PACE-BR (0.0292 sec)}, \texttt{PACE-RFR (0.0745 sec)}, and \texttt{PACE-MLP (0.1023 sec)} by (2x, 3x, 7x and 10x) respectively. Finally, on \texttt{DSD}, our primary predictor \texttt{PACE-kNN} attained a \texttt{($\mu$(SR, NR) = 0.0745 sec)} latency.

 \vspace{-2mm}
\begin{figure}[!htbp]
\centering
\captionsetup{justification=centering}
\subfloat[MSE ($\Delta$)]
{{\includegraphics[width=3.5cm]{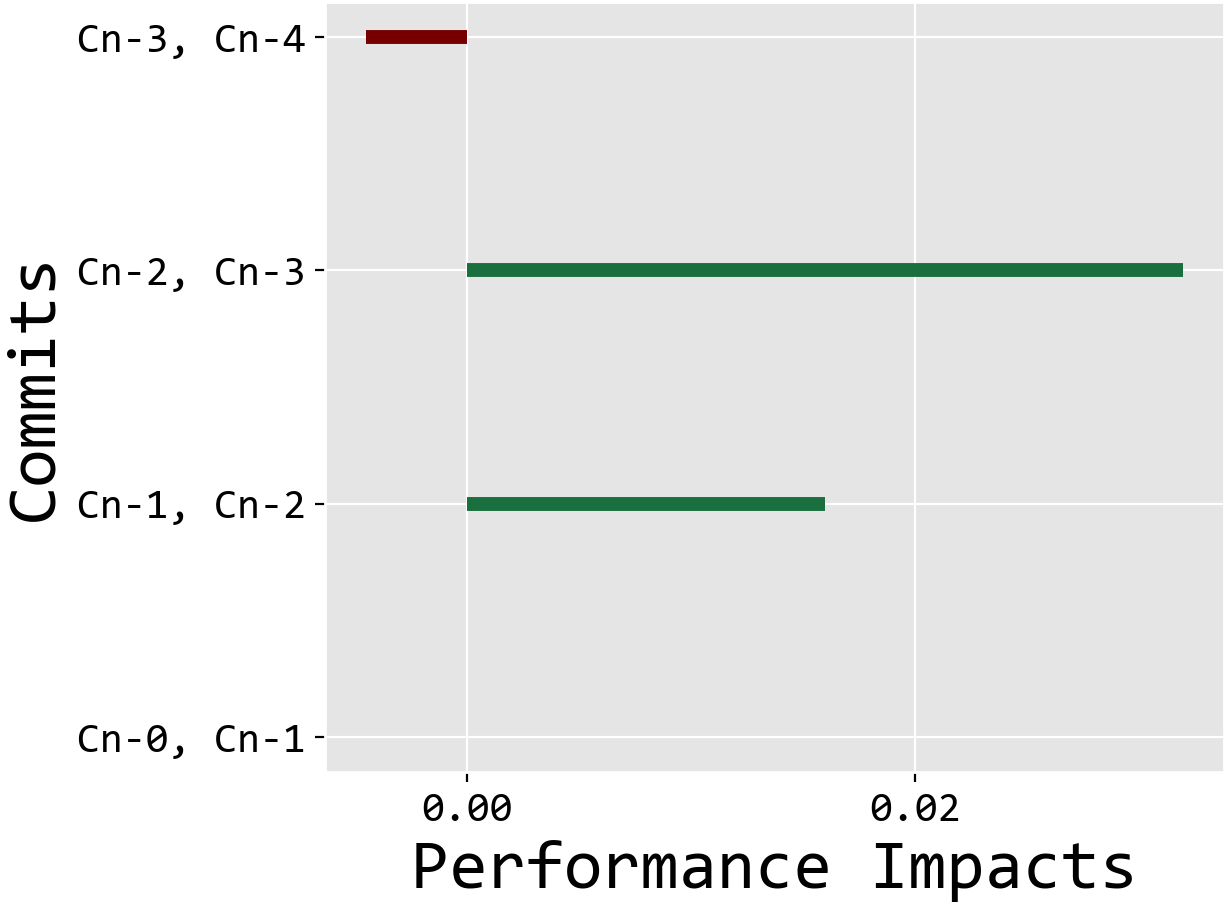}}}
\subfloat[MAE ($\Delta$)]
{{\includegraphics[width=3.5cm]{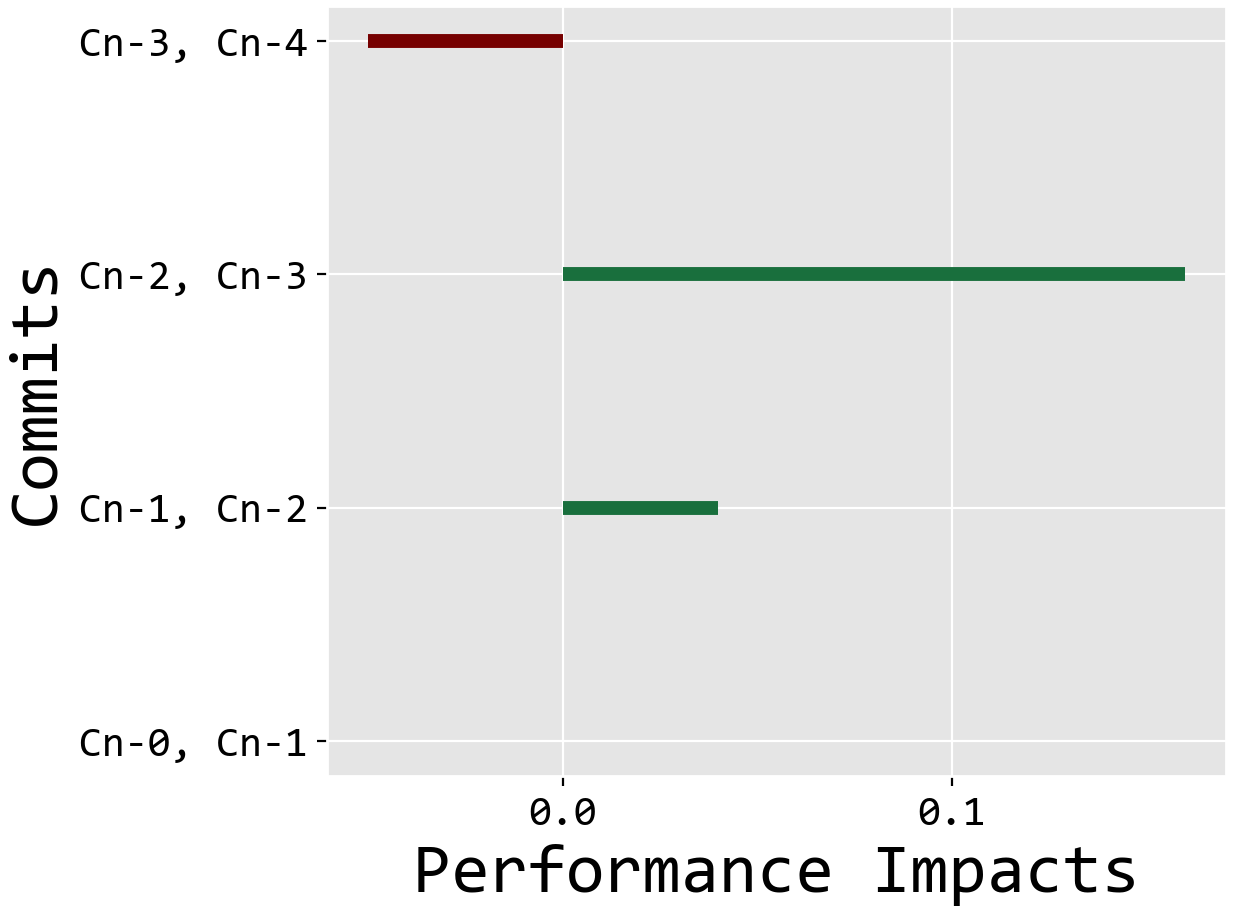}}}
\subfloat[RMSLE ($\Delta$)]
{{\includegraphics[width=3.5cm]{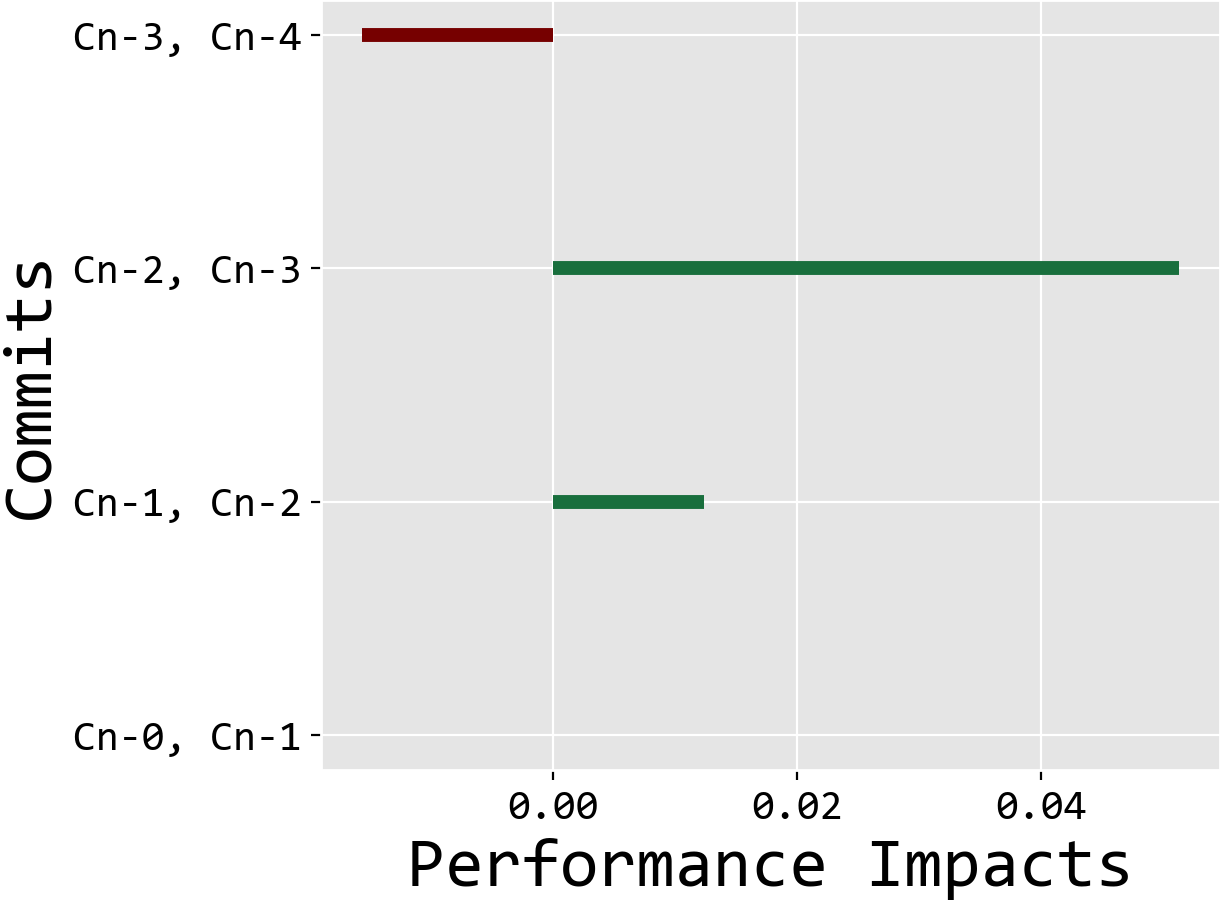}}}
\subfloat[AVG ($\Delta$)]
{{\includegraphics[width=3.5cm]{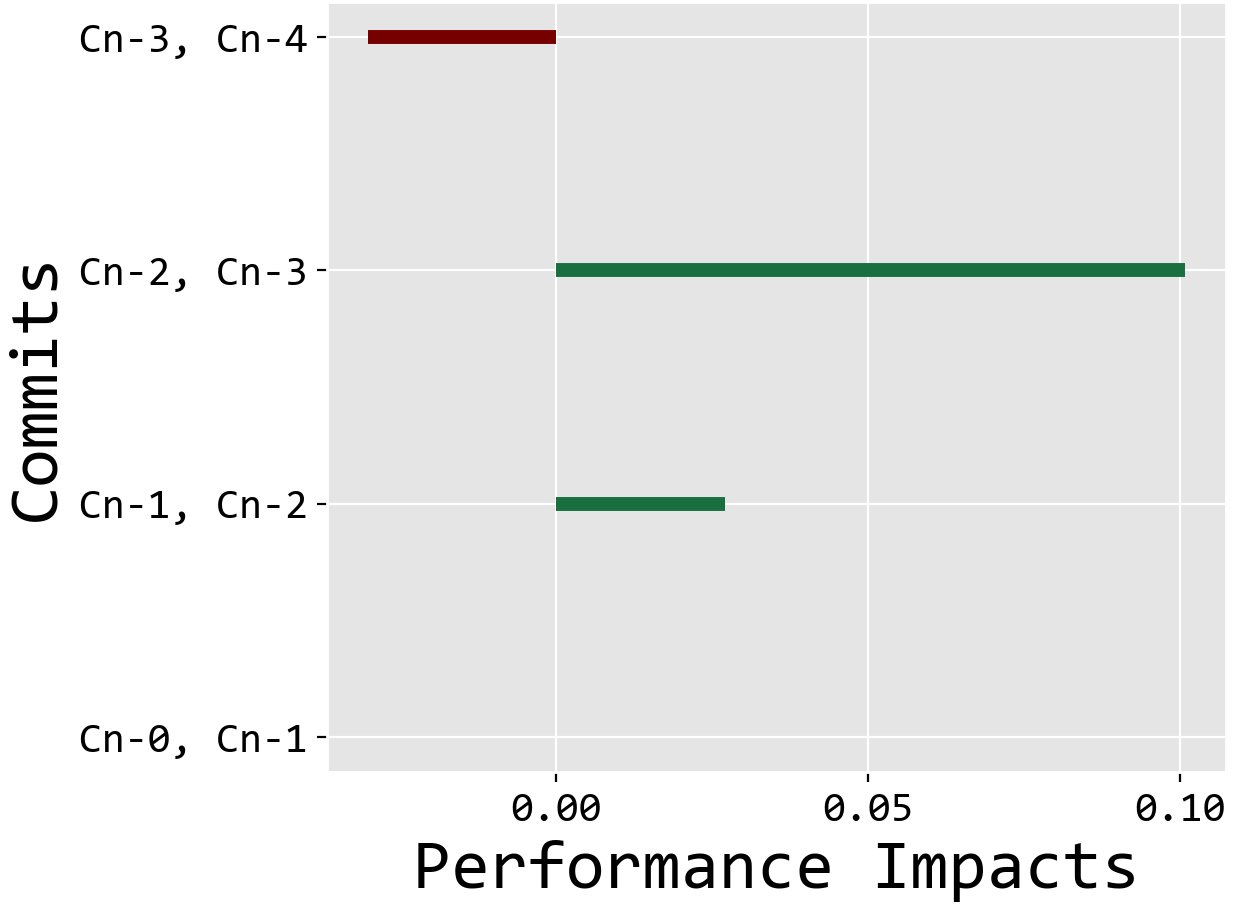}}}
\vspace{-2mm}
\caption{Performance Impacts of Code Update (ABD)}
\label{fig:delta_perf_impact}
\end{figure}

\vspace{-2mm}
\begin{figure}[!htbp]
\centering
\captionsetup{justification=centering}
\subfloat[MSE ($\Delta$)]
{{\includegraphics[width=3.5cm]{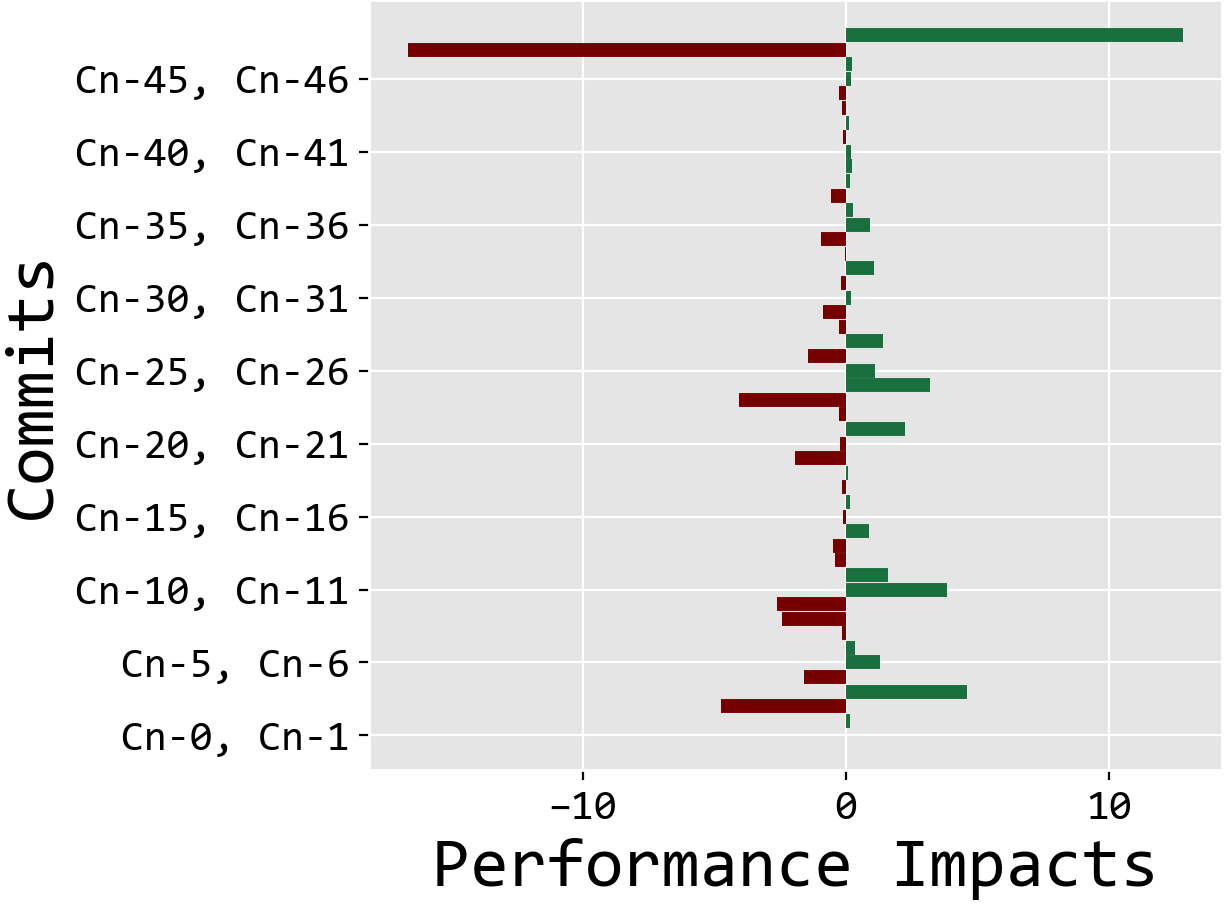}}}
\subfloat[MAE ($\Delta$)]
{{\includegraphics[width=3.5cm]{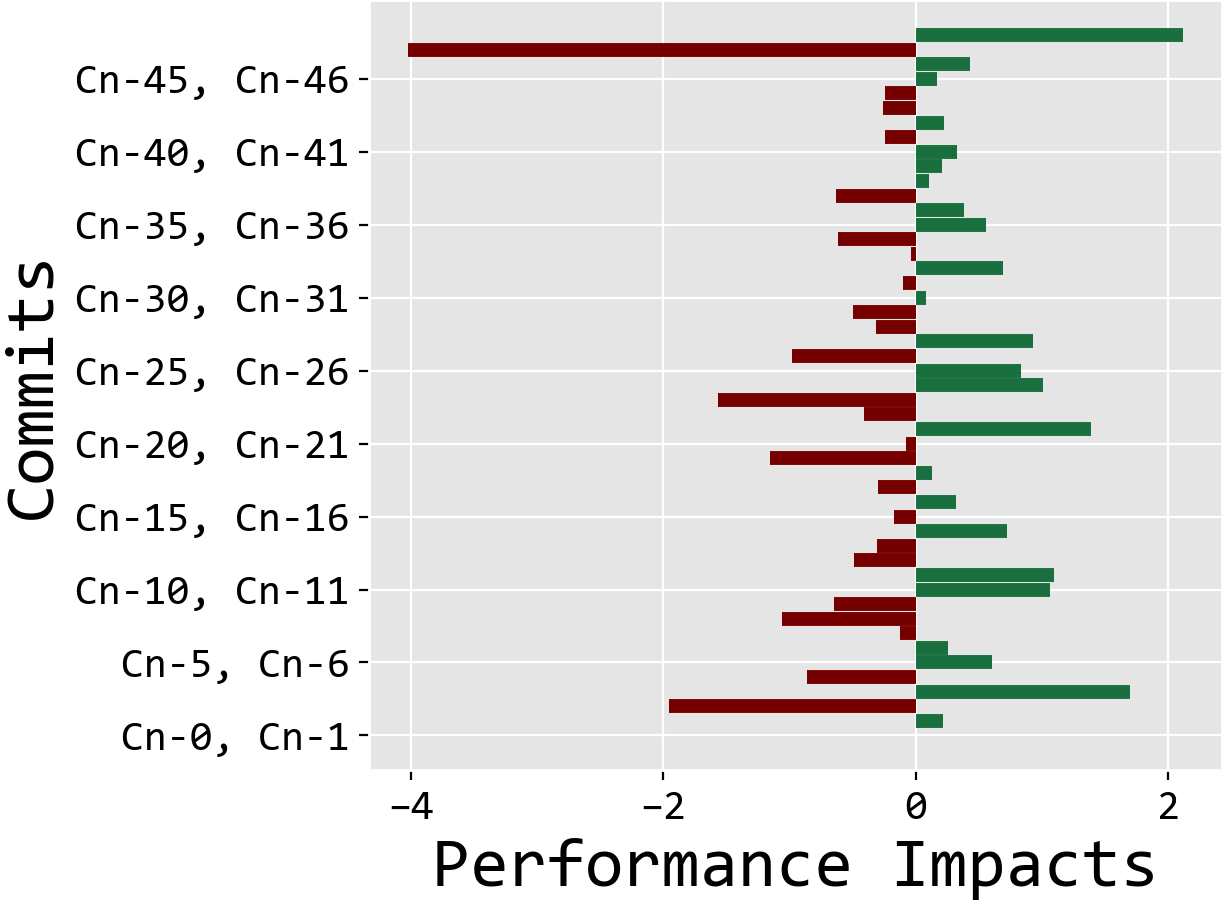}}}
\subfloat[RMSLE ($\Delta$)]
{{\includegraphics[width=3.5cm]{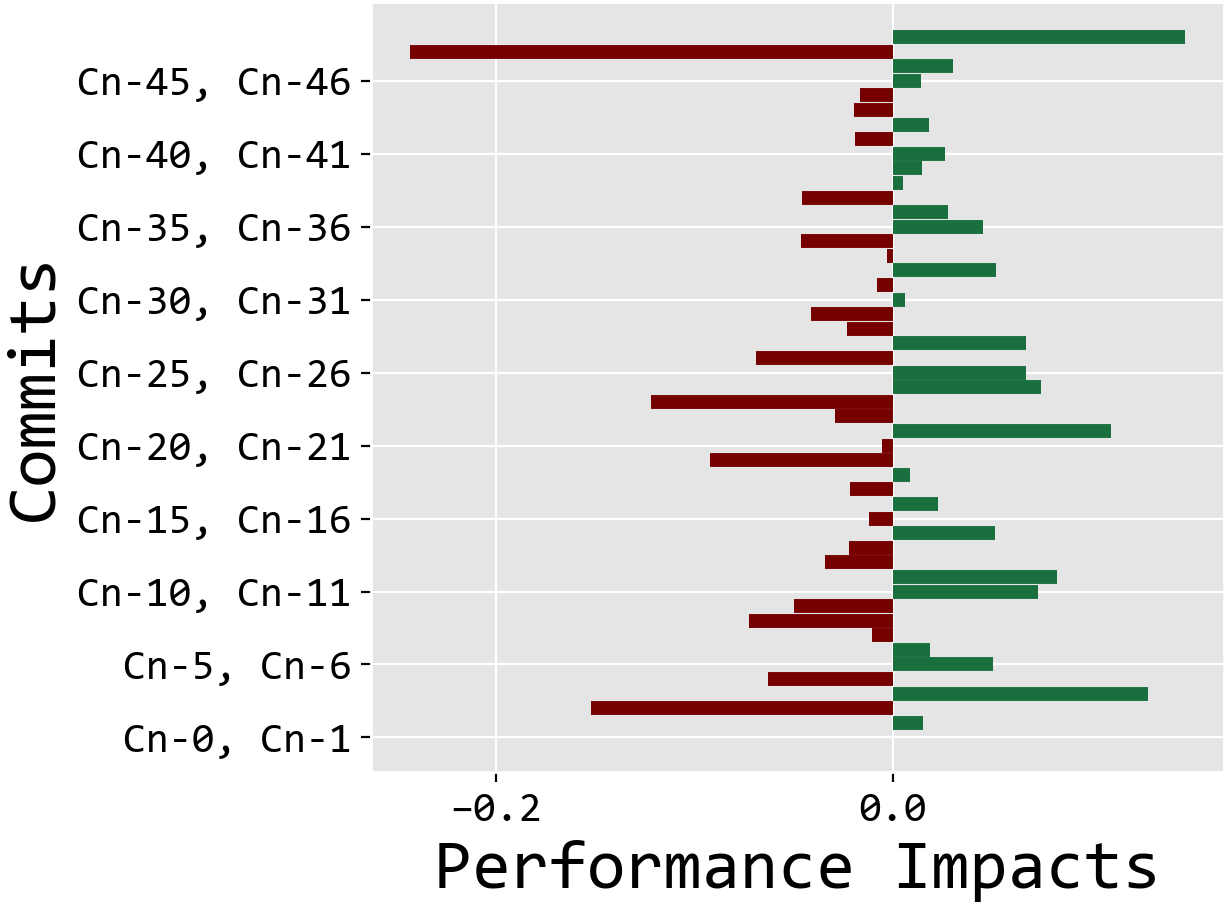}}}
\subfloat[AVG ($\Delta$)]
{{\includegraphics[width=3.5cm]{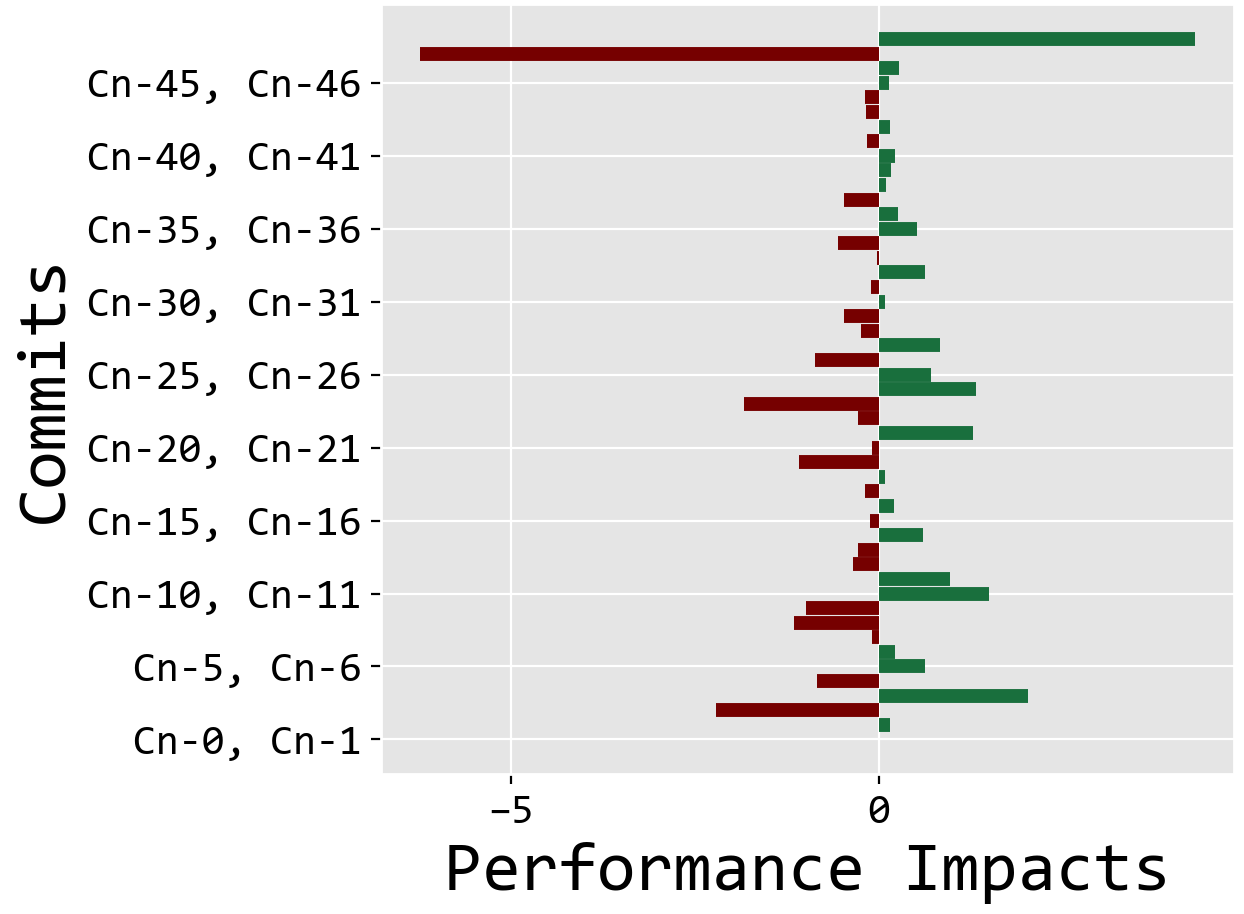}}}
\vspace{-2mm}
\caption{Performance Impacts of Code Update (DSD)}
\label{fig:delta_perf_impact1}
\end{figure}

\vspace{-2mm}

\vspace{-2mm}

\subsection{RQ2. What is the performance impact of commit at ($c\textsubscript{n-1}$) given ($c\textsubscript{n}$) code update?}

We answered this question by further analyzing the generated results from RQ1. We take the delta of the predictive performance of commit\textsubscript{i} given commit\textsubscript{i+1}. There are two categories (positive and negative) of performance impact. From Figures~\ref{fig:delta_perf_impact} and~\ref{fig:delta_perf_impact1}, commit\textsubscript{i} is considered to have a positive impact on the following commit\textsubscript{i+1}  if and only if the delta predictive error rate of commit\textsubscript{i+1}  is less than that of commit\textsubscript{i}, otherwise it is a negative performance impact. For instance, in Figure~\ref{fig:delta_perf_impact1}d, the average error rate for \texttt{DSD} is ($c\textsubscript{n}, c\textsubscript{n-1}$) is \texttt{0.2895}, the following update ($c\textsubscript{n-1}, c\textsubscript{n-2}$). c\textsubscript{n-1} is the training set fed to \texttt{PACE-kNN} to predict the performance c\textsubscript{n-2} (testing set), resulting in a \texttt{0.1399} average error rate. The delta is  \texttt{0.1496}. Hence, the performance impact between ($c\textsubscript{n}, c\textsubscript{n-1}$) and ($c\textsubscript{n-1}, c\textsubscript{n-2}$)  is positive. The average performance impact of $c\textsubscript{n-i}$ is (positive\texttt{=2}, negative\texttt{=1}) $\in$ \texttt{ABD} (Figure~\ref{fig:delta_perf_impact}d) and (positive\texttt{=24}, negative\texttt{=24}) $\in$ \texttt{DSD} for \texttt{3} and \texttt{48} updates respectively. On \texttt{ABD}, commits ($c\textsubscript{n-1}$, $c\textsubscript{n-2}$) and ($c\textsubscript{n-3}$, $c\textsubscript{n-4}$) has the best positive (\texttt{0.0270}) and single negative (\texttt{-0.0300}) $\Delta$ performance respectively. Similarly, \texttt{DSD}'s  ($c\textsubscript{n-18}$, $c\textsubscript{n-19}$) and ($c\textsubscript{n-47}$, $c\textsubscript{n-48}$) has the best positive (0.0862) and worst negative (-6.2362) $\Delta$ performance respectively.


\subsection{RQ3. What is the time cost of selecting and representing code stylometry features?}

\begin{figure}[!htbp]
    \centering.
    \captionsetup{justification=centering}
    \subfloat[ABD Feature Selection]
    {{\includegraphics[width=4.5cm]{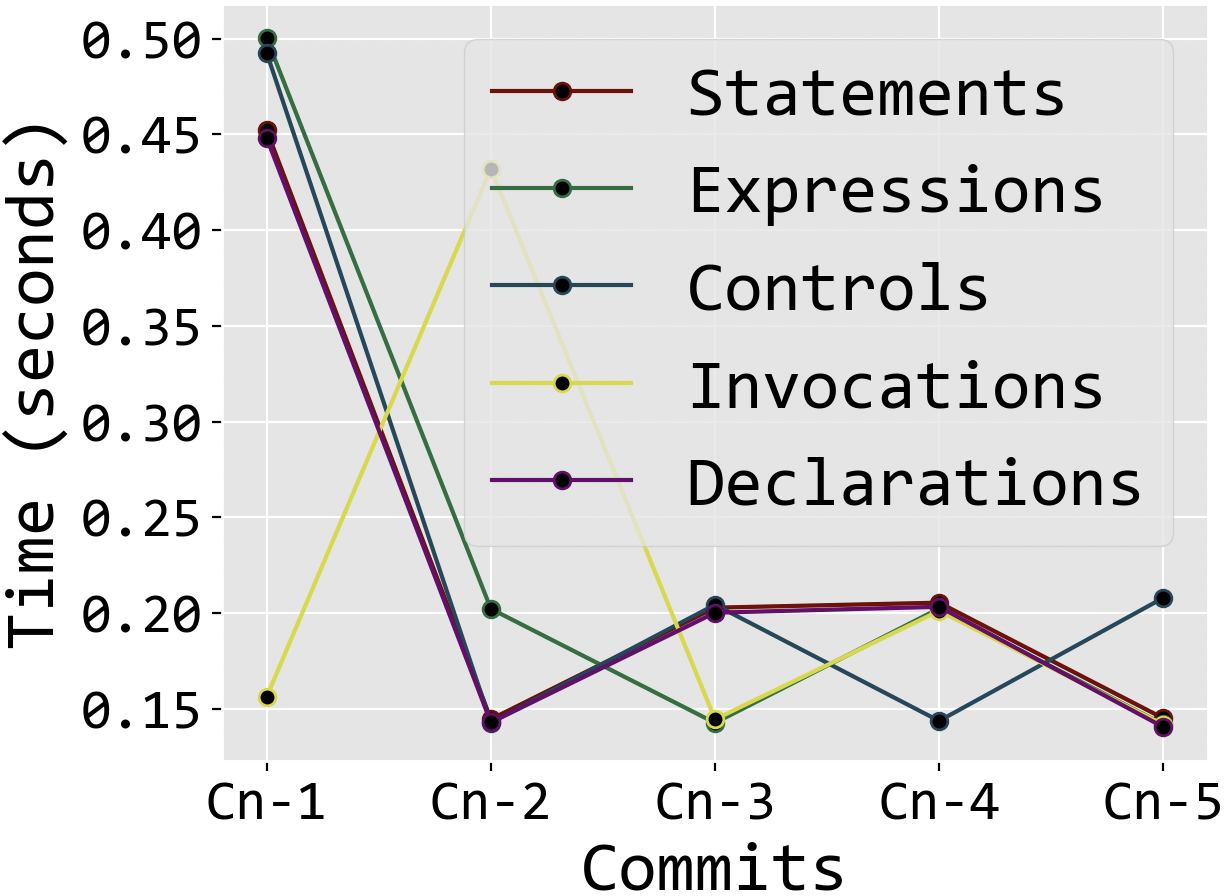}}}
    \subfloat[DSD Feature Selection]
    {{\includegraphics[width=4.5cm]{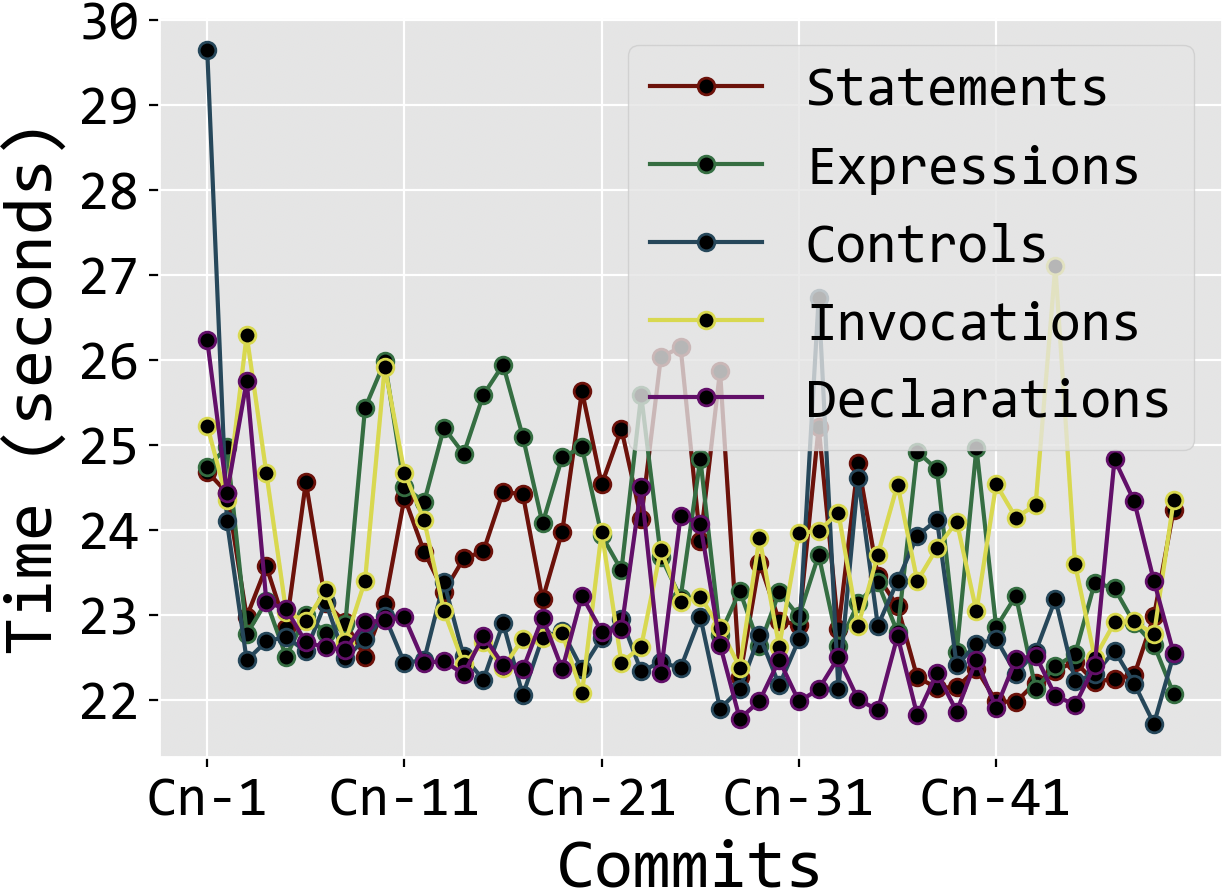}}}
    \subfloat[ABDSR]
    {{\includegraphics[width=4.5cm]{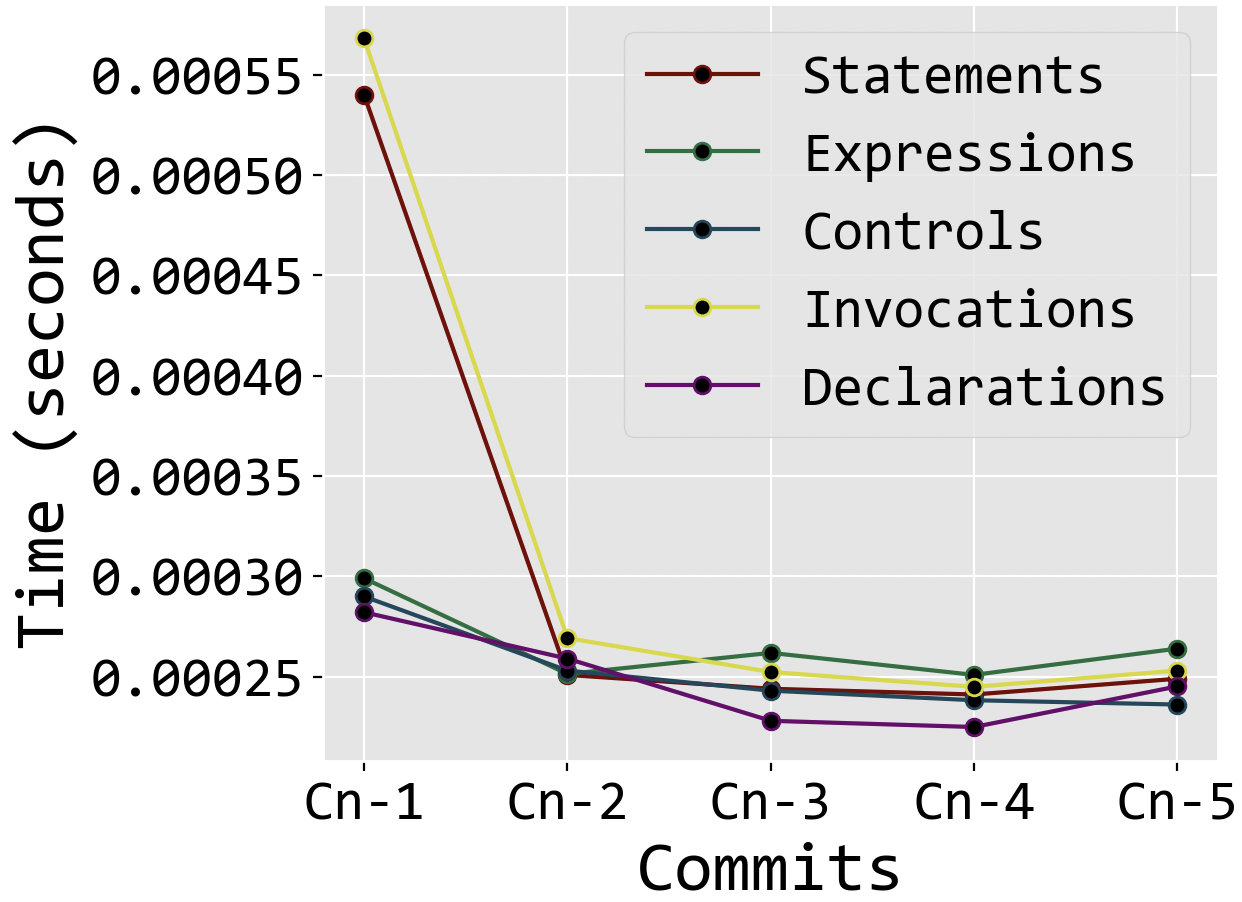}}}
    \vspace{-2mm}
    
    \caption{ABD and DSD CStyle Feature Selection and ABD Statistical Representation (ABDSR) \\
    Syntactic: $\mu$($\mu$(statement, expressions, controls)) = (0.2354, 23.3939) $\in$ (ABD, DSD)  \\
    Lexical: $\mu$($\mu$(invocations, declarations)) = (0.2211, 23.2094) $\in$ (ABD, DSD) \\
    $\mu$(Syntactic, Lexical) = (0.2297, 23.3201) $\in$ (ABD, DSD)   \\
    \texttt{ABDSR}: $\mu$($\mu$(statement, expressions, controls), $\mu$(invocations, declarations)) = 0.0002 \\    
    }
    
    \label{fig:ab_selection}
\end{figure}

\begin{figure}[!htbp]
    \centering.
    \captionsetup{justification=centering}
    \subfloat[DSSR]
    {{\includegraphics[width=4.5cm]{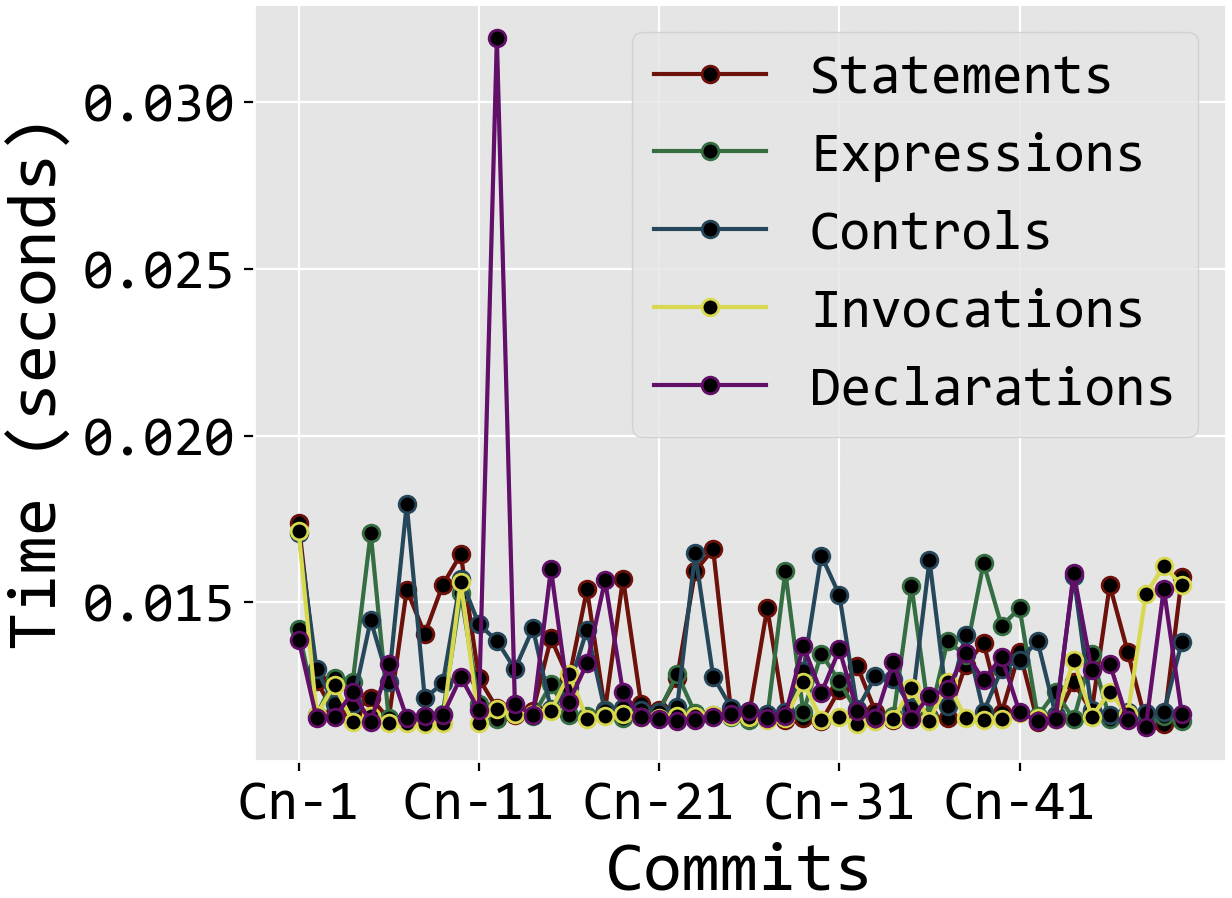}}}
    \subfloat[AB Neural Representation]
    {{\includegraphics[width=4.5cm]{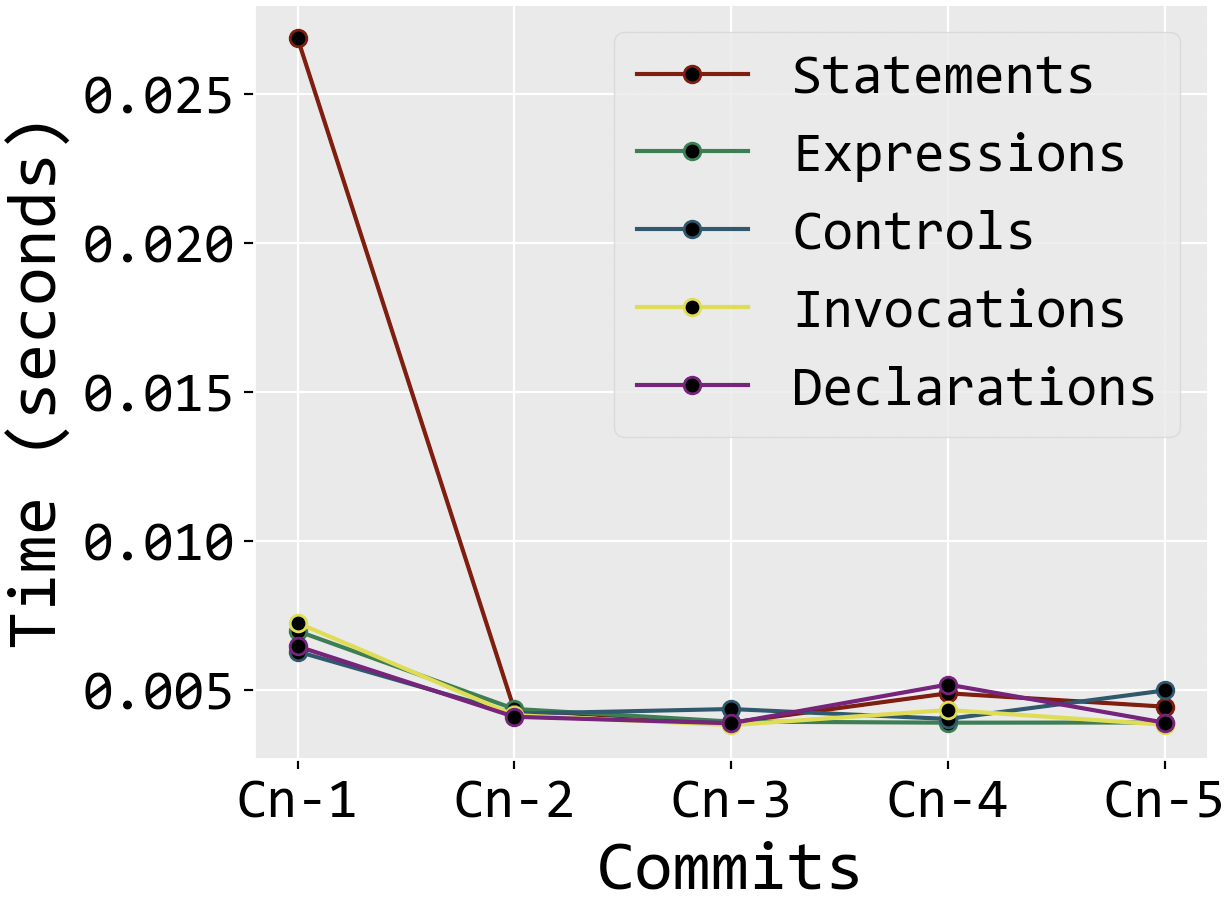}}}
    \subfloat[DS Neural Representation]
    {{\includegraphics[width=4.5cm]{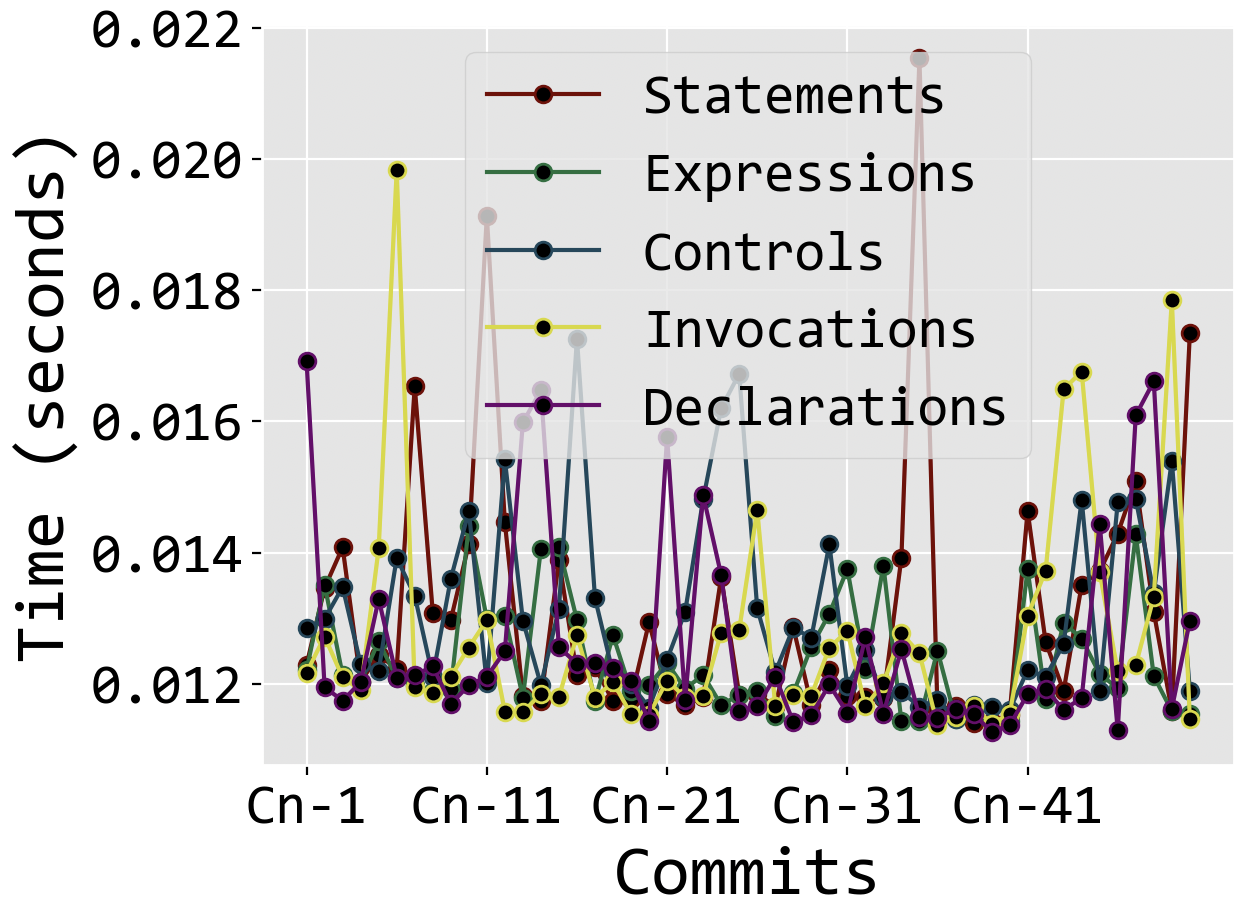}}}
    \vspace{-2mm}

    \caption{DSD Statistical Representation (DSDSR), ABD and DSD Neural Representations \\
    \texttt{DSDSR}: $\mu$($\mu$(statement, expressions, controls), $\mu$(invocations, declarations)) = 0.0055 \\ 
    Syntactic: $\mu$($\mu$(statement, expressions, controls)) = (0.2354, 23.3939) $\in$ (ABD, DSD)  \\
    Lexical: $\mu$($\mu$(invocations, declarations)) = (0.2211, 23.2094) $\in$ (ABD, DSD) \\
    $\mu$(Syntactic, Lexical) = (0.2297, 23.3201) $\in$ (AB, DS)   \\   
    }
    
    \label{fig:c_neural}
\end{figure}

We answer this question by timing the selection and representation of features for each CCS in the \texttt{ABD} and \texttt{DSD}. The \texttt{ABD} constitutes five commits and 42 average number of files (\texttt{ANF}), hence a much lower feature selection and representation throughput, while its DS counterpart has 50 commits and 2543 \texttt{ANF}. The $\mu$ cstyle (statements, expressions, controls, invocations, declarations) feature selection throughput for \texttt{AB} and \texttt{DS} datasets are \texttt{0.2297}$\mu$ sec and \texttt{23.3201}$\mu$ sec (see Figure~\ref{fig:ab_selection} a and b). After feature selection, we proceeded to time the statistic (\texttt{SR}) and neural representation (\texttt{NR}) of these features. SR is $\in$ (\texttt{0.0002}$\mu$ sec $\land$ \texttt{0.0126}$\mu$ sec) for \texttt{ABD} $\land$ \texttt{DSD}. \texttt{NR} is $\in$ (\texttt{0.0055} $\mu$ sec $\land$ \texttt{0.0588} $\mu$ sec) for \texttt{ABD} and \texttt{DSD} respectively (see Figures~\ref{fig:ab_selection}c, and ~\ref{fig:c_neural}). 

The primary objective of this RQ is to ascertain the throughput of cstyle feature selection and the fastest feature representation method between \texttt{SR} and \texttt{NR}. Thus, the statistical representation of cstyle features is 27x and 5x faster than its neural counterpart on \texttt{ABD} and \texttt{DSD} respectively. Note, the time difference in intra-comparative feature representations (syntactic against lexical $\in$ \texttt{SR} $\lor$ \texttt{NR}) is not attributed to the representation algorithm but to its frequency occurrence.

\vspace{-2mm}

\vspace{-2mm}

\subsection{RQ4. How does PACE perform in comparison with TEP-GNN?}
\label{subsec:rq3}

To the best of our knowledge, we have not encountered any work directly comparable to \texttt{PACE} with respect to scope, approach, and the mapping of statistical and neural represented cstyle features to execution test times as a performance microbenchmark for the continuous prediction code performance. However, we aimed to derive a robust evaluation of \texttt{PACE}. Hence we compare \texttt{PACE} to a closely related work in performance prediction titled \texttt{TEP-GNN}~\cite{samoaa2022tep}. \texttt{TEP-GNN} leverages the merging of flow-augmented AST (\texttt{FA-AST}) --- dynamic interpretation of ASTs~\cite{wang2020detecting} --- and Graphical Convolutional Neural Network (\texttt{GCNN}) -- \texttt{GCNN}'s are a subset of deep learning models that exploits graphically represented input observations to encode and transmit learned features using its vertices and edges~\cite{morris2019weisfeiler} -- to predict the execution time of functional test cases. 

The authors evaluated their work using four software repository case studies~\cite{samoaa2022tep}, (i) \texttt{SystemDS}: an open source ML system for conducting end-to-end data science tasks, (ii) \texttt{H2}: a Java-based relational database management system, \texttt{Dubbo}: a Java-based open source remote procedure call framework, and \texttt{RDF4J}: A Java-based resource description framework. The authors created their experimental datasets by adapting GitHub's hosted runners to build source files  ($x^{(i)}$) and consequently extract the execution times traces of published test results ($y^{(i)}$).

\begin{table}[ht]
\caption{\texttt{PACE-kNN's} performance in comparison with \texttt{TEP-GNN}. \#File: Total number of source files, \#Nodes: Total number of nodes, Combined: Entire dataset (H2, RDF4J, Dubbo, and SystemDS), \#VOC: Vocabulary size}
\resizebox{0.80\linewidth}{!}{
\begin{tabular}
{{ p{0.09\linewidth} p{0.08\linewidth} | p{0.08\linewidth} p{0.06\linewidth} p{0.08\linewidth} || p{0.08\linewidth} p{0.06\linewidth} p{0.18\linewidth} }}
\toprule
\multicolumn{4}{r}{\textbf{\texttt{TEP-GNN}}} 
&&
\multicolumn{3}{c}{\textbf{\texttt{PACE-kNN}}} \\
\toprule
\textbf{\texttt{Project}} & \textbf{\texttt{\#Files}} & \textbf{\texttt{\#Nodes}} & \textbf{\texttt{\#VOC}} & \textbf{\texttt{MSE}} & \textbf{\texttt{\#Nodes}} & \textbf{\texttt{\#VOC}} & \textbf{\texttt{MSE(SR,NR)}}
\\
\toprule
\texttt{H2} & 194 & 405706 & 17972 & 0.0340 & 113800 & 35 & (0.0056, 0.0119) \\  
\midrule
\texttt{RDF4J} & 478 & 214436 & 10755 & 0.0160 & 52004 & 30 & (0.0003, 0.0062) \\ 
\midrule
\texttt{Dubbo} & 123 & 75787 & 4499 & 0.0230 & 19304 & 31 & (0.0076, 0.0110) \\ 
\midrule
\texttt{SystemDS} & 127 & 110651 & 3161 & 0.0110 & 22803 & 30 & (0.0288, 0.0369) \\ 
\midrule
\texttt{Combined} & 992 & 806580 & 36387 & 0.0170 & 207911 & 126 & (0.0083, 0.0077) \\ 
\midrule
& & & \textbf{\texttt{AVG}} $\blacktriangleright$ & 0.0202 &  & \textbf{\texttt{AVG}} $\blacktriangleright$ & (0.0101, 0.0147) \\ 
 \bottomrule
\end{tabular}
}
\label{table:rq1}
\end{table}

We answer this question by replicating experimental steps adopted by the authors~\cite{samoaa2022tep}, which include: (i) retrieving the test files of case study datasets (\texttt{H2}, \texttt{RDF4J}, \texttt{Dubbo}, \texttt{SystemDS}) and corresponding execution run times, (ii) splitting the data into two sets, where 80\% $\in$ training and 20\% $\in$ testing, and (iv) evaluating predictions using MSE (see Section~\ref{subsec:eval_metrics}). \texttt{PACE-kNN} outperforms \texttt{TEP-GNN} on three (\texttt{H2}, \texttt{RDF4J}, and \texttt{Dubbo}) out of four projects by (143\%, 96\%), (192\%, 88\%), and (100\%, 70\%) \texttt{$\in$ (SR, NR)}. \texttt{TEP-GNN} outperformed \texttt{PACE-kNN} on \texttt{SystemDS} by (89\%, 108\%). Albeit, \texttt{TEP-GNN} best performance is on \texttt{SystemDS}, it also has the least linear correlation between actual and predicted values, as discussed by the authors. 

On average, \texttt{PACE-kNN} attained a (0.0083, 0.0077) error rate on the entire dataset and (0.0101, 0.0147) on the cumulative average performance. Hence, out-performing \texttt{TEP-GNN} by (68\%, 75\%) and (66\%, 31\%) $\in$ \texttt{(SR, NR)} respectively. Furthermore, \texttt{TEP-GNN} which was trained for 100 epochs has a significantly higher vocabulary size, where (17972, 10755, 4499, 3161, 4499, 3161) $\in$ (\texttt{H2, RDF4J, Dubbo, SystemDS}) compared to \texttt{PACE-kNN} (35, 30, 31, 30). Thus, \texttt{PACE-kNN}'s vocabulary size is lower \texttt{TEP-GNN} by 513x, 358x, 145x and 105x.

\section{Scope and Threats}
\label{sec:scope}
\subsection{Scope}
The execution time of functional test cases as a means of micro-benchmarking the code performance of software doesn't tell the whole story of what constitutes optimal performing software. Incessantly changing contributing factors such as disk space, memory, networks, operating systems, and core hardware renders a conclusive performance predictive model almost impossible and inherently a qualitative endeavor. However, it's viable to quantitatively solve the sub-components of the problem as demonstrated by our experiments. Hence, \texttt{PACE} narrows down constraints by assuming an ideal software scenario devoid of the volatility of memory, hardware compatibility, and network troubleshooting, which is outside the scope of this work. Furthermore, estimating the baseline execution time measure of automated test cases is valuable in specific use cases, such as non-collaborative development environments. However, it is not a substitute and outside the scope of \texttt{PACE}, as it defeats the purpose of predicting the performance impact of the code update to the primary before the commit process is complete, negating the introduction of unoptimized code to the primary repository in the first place.

\subsection{Threats}
\subsubsection{\texttt{\textbf{Availability}}.}
Detailed in Section~\ref{subsec:automated_testing}, the extraction execution times of functional tests are dependent on the availability of test cases in the software repository. These modules are responsible for testing the functionalities and features of the software to ensure they perform as anticipated. Hence, testing scripts should also be compact and have significant code coverage. Thus, in the absence of these scripts, devising a technique to extract execution test times whilst maintaining our principle of code correctness is pertinent and subject of future research. 

\subsubsection{\texttt{\textbf{Validity}}.} 
Albeit diligently selecting candidate repositories that satisfy requirements for a professional software repository, it is still possible to have projects where test cases are sparse in breadth and shallow in depth, constituting a false positive, which would be problematic for code correctness. One pathway to approach this problem would be to design a \textit{\texttt{search+validate}} system to search and validate the existence or lack thereof of functional test cases. For example, given a program with modules, there exist routines, search for associated test cases to validate routine correctness, assign a positive value (\texttt{+ve}), otherwise assign a negative value (\texttt{-ve}). Consequentially, it would be up to the users to determine what constitutes a permissible threshold value or is reflective of the tests' depth and breadth. 

\subsubsection{\texttt{\textbf{Generalizability}}.}
Given case studies projects written in Java, an apparent cause for concern is  \texttt{PACE's} generalizability. This work doesn't empirically demonstrate that the entirety of \texttt{PACE} methods is transferable to other software projects. However, we argue that candidate repository selection, code stylometry feature engineering, and functional testing components of \texttt{PACE} can be generalized. The candidate repository selection process was to find repositories that may be representative of large software development projects. Microbenchmarking duty is with the CI infrastructure, not with the code or software developer, and we do provide the open-source repository for that. Any CI infrastructure can use the same method and apply our proposed methods.

The selection and representation of cstyle features are foundational to the imperative programming paradigm. For example, the existence of an \texttt{if}, \texttt{while} statement conditionals and \texttt{lambda} expressions in Java are represented with (\textit{IfStatement, WhileStatement, LambdaExpression}) nodes. Similarly, the aforementioned features nodes in Python are represented with (\textit{If}, \textit{While}, \textit{Lambda}) nodes using the ast module of the Python~\footnote{\texttt{https://docs.python.org/3/library/ast.html}} standard library. Thus, the concern of cstyle feature engineering generalizability can be tackled by simply matching feature nodes in this work to corresponding nodes in other high-level imperative programming languages.

\subsubsection{\texttt{\textbf{Outliers}}.}
Another obvious threat is in the area of \textit{disproportionate code updates}, which are voluminous introductions or deduction of source files to the software that skew the baseline test performance. Recall our worst predictive performance \texttt{($\mu$(RMMR) = 6.2660)} on \texttt{DSD} occurred at commits (\texttt{$c_{n-48}, c_{n-49}$}) and had the most file and test times difference between commits. Given that \texttt{PACE} has no control over the content of the code base, we argue that the best way to solve this threat is to integrate an observer program that listens and notifies users of voluminous updates, enabling the user to be alert to potentially outlier values when evaluating corresponding predictions. Furthermore, as demonstrated in RQ1, \texttt{PACE-kNN} model learns the new features resulting in improved predictions in subsequent commits.

\subsubsection{\texttt{\textbf{Base Case (What is a good starting n?)}}.}
Recall that to simulate a pseudo-active software development scenario, we introduced \texttt{RPiTs} (see Section~\ref{subsec:data_collection}, paragraph 3). \texttt{RPiTs} reverse development time to freeze and extract the current code state given a (\texttt{n}) which enables the retrieval of software state throughout its development history. Our case study datasets \texttt{ABD} and \texttt{DSD} have 5 and 50 commit n's respectively. Given that developing software commonly commences with a developer writing a character, a collection of characters (string), a collection of strings (function) a collection of functions (module), a collection module (package), a collection of packages (library), it is pertinent that \texttt{PACE}'s base case (n) is a collection of strings (function) at the minimum for the source code to be compiled, AST generated and cstyle features extracted. Following that, it becomes apparent that a collection of characters (string) won't suffice because (i) not enough observations for predictive modeling and (ii) complexity in designing in writing functional test cases to validate tiny components of a code. Thus, for these reasons, we ensured that our \texttt{RPiTs} technique reflects these considerations. Hence, a threat constituting source files usually occurring in the beginning phases of developing the software becomes evident. We argue that there are two avenues to approach this threat. 

The first entails a modification to \texttt{RPiTs} to ensure that the base case for \texttt{n} always has a non-trivial ($\geq$ 10) number of routines with its corresponding test cases suitable for functional testing and predictive modeling. Second, exploiting the success of large language models (LLMs)~\cite{zhao2023survey} --- contextually distributional semantic models trained using the Transformer neural architecture on colossal amounts of data to perform a specific task, such as representative (\texttt{Bidirectional Encoder Representations from Transformers (BERT)}), generative (\texttt{Generative Pre-Trained Transformer}), and pseudo-universal (\texttt{Pathways Language Model (PaLM)}) ---  in several language tasks to fine-tune limited \texttt{n} observations, enabling it to benefit of using pre-trained weights to predict associated test times.

\section{Related Work}
\label{sec:related_work}
Software performance predictions are an inherently complex task akin to building optimally compatible software with hardware machines of varying types, operating systems, and network infrastructure. Given the breakneck speed of technological advancement, it is a challenge to consistently account for these variables while maintaining highly accurate performance predictions. However, recent advances in the literature have applied machine and deep learning methods to achieve this goal.

Research work in the literature with varying degrees of predictive performance includes, Samoaa~\textit{et al.}~\cite{samoaa2022tep} proposed an approach for precise execution time prediction of functional tests using GNNs (see RQ4). Didona~\textit{et al.}~\cite{didona2015enhancing} combine analytical modeling and machine learning to magnify the traceability and overall applicability of performance prediction on NoSQL and Order Broadcast services. Meng \textit{et al.}~\cite{meng2017mira} proposed a framework for creating performance models through static code targeted at the performance of scientific applications. Guo \textit{et al.}~\cite{guo2018data} proposed a data-efficient approach to predict the performance of configurable systems. 

Zhou~\textit{et al.}~\cite{zhou2019deeptle} proposed a classification model based on the extraction of semantic and structural features to predict the performance (Accepted and Time Limit Exceed) of competitive programming solutions before submission. Kaltenecker \textit{et al.}~\cite{kaltenecker2020interplay} discusses the tradeoffs in employing ML for performance prediction. The authors recommended adopting a more nuanced downstream approach appropriate for the targeted problem domains. Ramadan~\textit{et al.}~\cite{ramadan2021comparative} learned the representation of hierarchical structural changes of a source code used to build a Long Short-Term Memory (LSTM) model tasked with code performance predictions. 

 Liu~\textit{et al.}~\cite{liu2021using} proposed a max/min algorithm to predict the code execution time performance of digital signal processing (DSP) software. The proposed algorithm is a fusion of clustering, similarity measure, application of weights in sample and attribute, and linear regression. The authors validated their model using PHY DSP Benchmark on the TIC64 DSP processor obtaining a $\mathtt{\sim}$10\% error rate. Velez~\textit{et al.}~\cite{velez2021white} proposed Comprex, a system designed to have the explainable capacity to be associated with the performance of a configurable system. The authors evaluated its white-box system on open-source projects achieving similar performance to its black-box counterpart with a reduction in cost and an increase in ML explainability. 

Samoaa~\textit{et al.}~\cite{samoaa2023unified} proposed a unified active learning framework for annotating graphically represented data with applicability in predicting software performance. This work was motivated by the limited availability of annotated software performance graphical data required for training machine learning models. Their approach consists of representing source code using ASTs and transforming these trees to graph embeddings using FA-AST, supervised and unsupervised learning. Finally, the authors employ active and passive learning to query and annotate data. 

B{\"o}ck~\textit{et al.}~\cite{bock2023performance} conducting experiments to empirically demonstrate the utility of large language models (LLMs) in predicting the runtime performance of programming competition programs. They demonstrate that although these models produce accurate code performance predictions, they fail to generalize to out-of-sample domains such as video games. In their approach, they mined and fed the Code4Bench dataset to an AST-based neural network for feature learning. These features are used as inputs to classification and regression models to predict a program's execution time, program ID and associated runtime, and comparative predictions on program pairs. Experimental results show that although they achieved high accuracy in execution time prediction, the LLMs failed to generalize on unseen data when fine-tuned on \textit{Ubisoft} production code. 

\section{Conclusion}
\label{sec:conclusion}
Although training developers in writing high-performance code could be a viable solution, it is outside the scope of our work. Education quality and level of developers, training opportunities, and willingness to provide at organizations vary. Furthermore, training takes up significant financial and human resources while simultaneously prone to human-caused error. Thus, this paper presented \texttt{PACE}, a program analysis framework that learns and maps program features the test time microbenchmarks for continuous performance prediction.

\section{Acknowledgements}
This research was supported in part by UMass Dartmouth's Marine and Undersea Technology (MUST) Research Program funded by the Office of Naval Research (ONR) under Grant No. N00014-23-1–2141. The views and conclusions expressed in this paper are those of the authors and do not reflect the official policy or position of the University of Massachusetts Dartmouth, the Office of Naval Research, U.S. Navy, U.S. Department of Defense, or U.S. Government.

\bibliographystyle{ACM-Reference-Format}
\bibliography{reference}
\end{document}